\documentclass[reprint,
superscriptaddress,
 amsmath,amssymb,
 aps,
]{revtex4-2}

\usepackage{graphicx}
\usepackage{dcolumn}
\usepackage{bm}
\usepackage{hyperref}

\usepackage{xcolor}

\begin{document}

\preprint{APS/123-QED}

\title{Strong coupling between a topological insulator and a III-V heterostructure at terahertz frequency}

\author{D. Quang To}%
\affiliation{Department of Materials Science and Engineering, University of Delaware, Newark, DE 19716, USA}%

\author{Zhengtianye Wang}
\affiliation{Department of Materials Science and Engineering, University of Delaware, Newark, DE 19716, USA}%

\author{Q. Dai Ho}%
\affiliation{Department of Materials Science and Engineering, University of Delaware, Newark, DE 19716, USA}%

\author{Ruiqi Hu}%
\affiliation{Department of Materials Science and Engineering, University of Delaware, Newark, DE 19716, USA}%

\author{Wilder Acuna}
\affiliation{Department of Materials Science and Engineering, University of Delaware, Newark, DE 19716, USA}%

\author{Yongchen Liu}
\affiliation{Department of Materials Science and Engineering, University of Delaware, Newark, DE 19716, USA}%

\author{Garnett W. Bryant}%
\affiliation{Nanoscale Device Characterization Division, National Institute of Standards and Technology,
Gaithersburg, MD 20899-8423, USA}%

\author{Anderson Janotti}%
\affiliation{Department of Materials Science and Engineering, University of Delaware, Newark, DE 19716, USA}%

\author{Joshua M.O. Zide}%
\affiliation{Department of Materials Science and Engineering, University of Delaware, Newark, DE 19716, USA}%

\author{Stephanie Law}%
 \email{slaw@udel.edu}
\affiliation{Department of Materials Science and Engineering, University of Delaware, Newark, DE 19716, USA}%

\author{Matthew F. Doty}%
 \email{doty@udel.edu}
\affiliation{Department of Materials Science and Engineering, University of Delaware, Newark, DE 19716, USA}%


\date{\today}

\begin{abstract}
We probe theoretically the emergence of strong coupling in a system consisting of a topological insulator (TI) and a III-V heterostructure using a numerical approach based on the scattering matrix formalism. Specifically, we investigate the interactions between terahertz excitations in a structure composed of Bi$_{2}$Se$_{3}$ and GaAs materials. We find that the interaction between the Bi$_{2}$Se$_{3}$ layer and AlGaAs/GaAs quantum wells with intersubband transitions (ISBTs) in the terahertz frequency regime creates new hybrid modes, namely Dirac plasmon-phonon-ISBT polaritons. The formation of these hybrid modes results in anti-crossings (spectral mode splitting) whose magnitude is an indication of the strength of the coupling. By varying the structural parameters of the constituent materials, our numerical calculations reveal that the magnitude of splitting depends strongly on the doping level and the scattering rate in the AlGaAs/GaAs quantum wells, as well as on the thickness of the GaAs spacer layer that separates the quantum-well structure from the TI layer. Our results reveal the material and device parameters required to obtain experimentally-observable signatures of strong coupling.  Our model includes the contribution of an extra two-dimensional hole gas (2DHG) that is predicted to arise at the Bi$_{2}$Se$_{3}$/GaAs interface, based on density functional theory (DFT) calculations that explicitly account for details of the atomic terminations at the interface. The presence of this massive 2DHG at the TI/III-V interface shifts the dispersion of the Dirac plasmon-ISBT polaritons to higher frequencies. The damping rate at this interface, in contrast, compensates the effect of the 2DHG. Finally, we observe that the phonon resonances in the TI layer are crucial to the coupling between the THz excitations in the TI and III-V materials.  
\end{abstract}

\maketitle

\section{Introduction}

In the past decade, hybrid quantum materials have received remarkable attention as an emerging class of materials and promising candidates for applications ranging from conventional opto-electronics and spintronics to quantum computing, quantum simulation, and quantum communication\cite{Rejab2020,Liu2019,Liao2019,Hashtroudi2020}. In the simplest picture, one could simultaneously take advantage of the properties of more than one of the individual constituent materials that comprise the hybrid. More exciting is the possibility for such hybrids to have emergent properties that enable new electrical or optical functionality that exceeds what is possible in any individual material constituent. In particular, hybrid materials offer the opportunity to study and control the light-matter interactions that provide the foundation for important technologies such as X-ray sources, spectroscopy, biosensing, quantum information processing, and lasers \cite{Rivera2020,Torma2014,Zhang2012}. Specifically, the proximate environment can significantly alter the optical properties of emitters when the interaction between the emitters and the environment is sufficiently strong \cite{Zhang2012,Torma2014}. As a result, significant effort has been devoted to maximizing light-matter interaction in order to reach the strong coupling, or even ultrastrong coupling, regime in optical systems \cite{Diaz2019, Kockum2019}.

Hybrid materials are particularly intriguing for applications in the THz frequency band, a spectral window in which there are substantial opportunities for applications in medical diagnostics, security screening, bio-agent detection, telecommunication, high-speed electronics, and industrial quality control. A significant obstacle to the development of THz device technologies is the dearth of crucial components including high-power broadband photon sources, waveguides, modulators, gates, and detectors. Development of such components is hampered by the limitations of available materials \cite{Ferguson2002,Tonouchi2007,Wu2021}. For example, III-V semiconductors can generate and detect THz photons, but it is difficult to use them to handle other essential functions such as manipulating the THz polarization, guiding THz photons on-chip, or inducing nonlinear interactions that could be the foundation of logical gates. Likewise, topological insulators (TIs) can propagate THz plasmon polaritons, but cannot directly generate or detect them. A hybrid material containing both a TI and a III-V semiconductor could provide a material platform that overcomes these obstacles and enables optoelectronic device applications in the THz frequency domain. There have been many investigations of the underlying physics of Dirac plasmon polariton (DPP) excitations in a topological insulator materials \cite{Pietro2013,Politano2015,Kogar2015,Zhao2015,Autore2015,Politano2017,Ginley2018,Wang2020,Pietro2020} and the intersubband transition (ISBT) in III-V heterostructures \cite{Zaluzny1998,Plumridge2008,Zaluzny2013,Shekhar2014,Bondarenko2015,Zaluzny2018}. However, there has not been a comprehensive examination of the coupling between a TI and a III-V heterostructure and the device and material parameters required to create hybridized states or reach the strong coupling regime.

Hybridized states form when two distinct excitations interact with sufficient strength to create a new mode whose character and dispersion relation cannot be understood by considering either excitation alone. A good example is the formation of a surface plasmon polariton which is a hybridized state formed from an electromagnetic wave (photon) and charges oscillating at a metallic sample surface (plasmon). The emergence of such a hybridized state is typically observed through an anti-crossing (avoided crossing) in the dispersion relation. By analogy to cavity quantum electrodynamics, we define strong coupling to be the regime in which the observed mode splitting $(2g)$ becomes comparable to or larger than twice the linewidth of either involved excitation, i.~e.~$g \gtrsim \gamma_{TI}, \gamma_{ISBT}$ where $\gamma_{TI}$ ($\gamma_{ISBT}$) are the linewidths of the DPP in the TI (ISBT of the III-V Quantum Wells). These linewidths originate in the loss (dissipation) for each excitation and $\gamma_{TI}$, in particular, is strongly influenced by the interface between the TI and III-V material. The recent report of the successful fabrication of an atomically-clean and precise interface between MBE-grown Bi$_{2}$Se$_{3}$ and GaAs materials \cite{Liu2022} now makes it feasible to explore strong coupling between the DPPs in a TI layer and the collective ISBT excitations in a III-V heterostructure. 

\begin{figure}[h]
\centering
    \includegraphics[width=.5\textwidth]{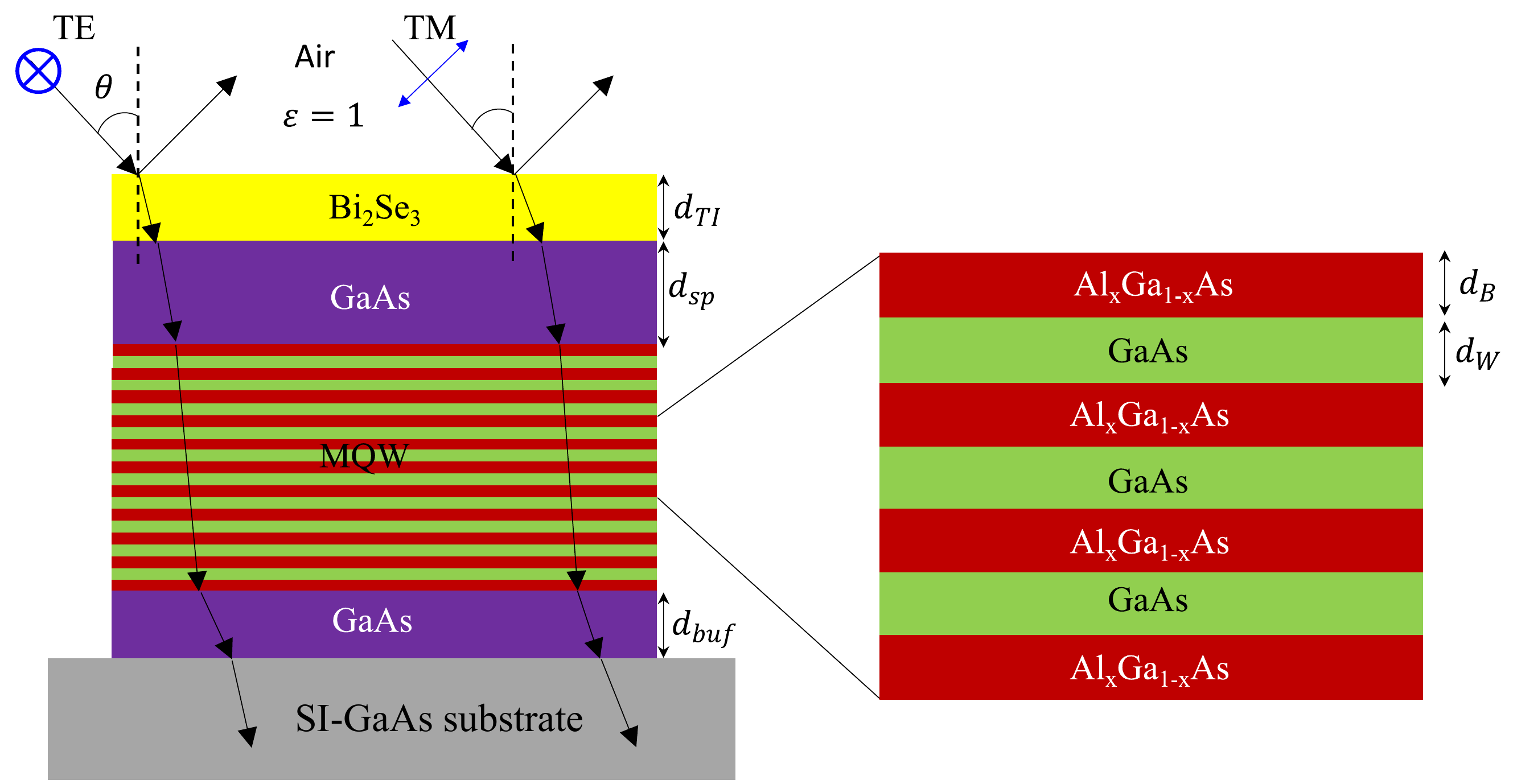}
 \caption{Schematic of the TI/III-V heterostructure comprising a Bi$_{2}$Se$_{3}$ layer and a single or multiple AlGaAs/GaAs quantum wells separated by GaAs spacer layer with a SI-GaAs substrate and a GaAs buffer layer to be considered in this work. The thickness of each layer is indicated in the figure with a typical thickness of GaAs well chosen to obtain the resonance of the intersubband transition at $6~THz$.}
  \label{FIG1}
\end{figure}

In this paper, we examine theoretically the coupling between the DPPs on the surface of a topological insulator layer and the ISBT in III-V quantum-well structures. We analyze how the formation of hybridized states, and the corresponding anti-crossing, depend on device and material quality parameters and thus reveal what must be achieved experimentally to reach the strong coupling regime as defined above. The device geometry we consider is shown in Fig.~\ref{FIG1}: starting from a semi-insulating GaAs substrate we imagine deposition of 1) a GaAs buffer layer, 2) a sequence of AlGaAs/GaAs quantum wells with ISBT transitions in the $1-9~THz$ frequency range, 3) a GaAs spacer layer of controlled thickness, and 4) a Bi$_{2}$Se$_{3}$ layer. We model this device using a semiclassical approach in which we solve Maxwell's equations to derive a dispersion relationship for the plasmon polariton in the Bi$_{2}$Se$_{3}$ layer as a function of varying interactions with the ISBTs, controlled by both the number of AlGaAs/GaAs quantum wells and the thickness of the GaAs spacer layer. Specifically, our calculations are based on a scattering matrix (or equivalently transfer matrix) method that provides a powerful tool for dealing with layered structures \cite{Wang2020}. We also performed density functional theory calculations to explore the emerging states at the Bi$_{2}$Se$_{3}$/GaAs interface. Our DFT result reveals an extra two dimensional hole gas (2DHG) occurring at the Bi$_{2}$Se$_{3}$/GaAs interface due to Se-terminated GaAs(001) at this interface. 
This 2DHG is included in our scattering matrix approach as an additional term for the optical conductivity of the DPP in the pristine Bi$_{2}$Se$_{3}$ layer, which is taken into account in the boundary conditions of Maxwell's equations when solving using our approach. 

The paper is structured as follows. In Sect.~\ref{method}, we present the methods and models employed in this paper to investigate the coupling between the Bi$_{2}$Se$_{3}$ layer and the multiple AlGaAs/GaAs quantum wells separated by the GaAs spacer layer. We first review the DPP system on the surface of a pristine Bi$_{2}$Se$_{3}$ layer (Sect.~\ref{DiracBiSe}) and the ISBT in a single AlGaAs/GaAs quantum well (Sect.~\ref{III_V_ISBT}) to obtain the separate optical responses to the electric and magnetic components of an electromagnetic wave propagating within each constituent material. We then describe our density functional theory-based calculation to obtain the electronic band structure of the Bi$_{2}$Se$_{3}$/GaAs interface (Sect.~\ref{DFTpredict}). This calculation predicts the existence of a 2DHG at the Bi$_{2}$Se$_{3}$/GaAs interface that is distinct from the topological surface states of the pristine Bi$_{2}$Se$_{3}$ layer. Sect.~\ref{method} concludes with a description of the scattering matrix method we use to solve Maxwell's equations within the layered structure of the devices we consider (Sect.~\ref{ScatteringMethod}). In Sect.~\ref{results} we discuss the calculated dispersion relations for the surface plasmon-phonon-ISBT polaritons. We explore the dependence of these dispersion relations on various material properties and, in particular, explore the material and device properties required to obtain strong coupling between the TI and the III-V heterostructure. The roles of the material parameters in tuning the strength of this coupling provide an important guidance as to how the strong coupling regime can be reached in experimentally practical devices. Finally, we give conclusions and perspectives for this work in Sect.~\ref{conc}.

\vspace{5mm}

\section{Methods and models}
\label{method}

Our goal is to understand the coupling between the DPP in a TI and the ISBT in a III-V heterostructure, and to explore the material and structural parameters that enable us to reach the strong coupling regime. Our primary tool is analysis of the dispersion relationship (energy vs wave vector, $E(k)$, or equivalently frequency vs wave vector, $\omega (k)$) of an electromagnetic wave propagating in the system as a function of structural and material parameters. We generate the dispersion relations by using the scattering matrix method to solve Maxwell's equations subject to the standard electromagnetic boundary conditions at the interfaces between the constituent materials. From the output of these scattering matrix calculations we plot the imaginary part of the reflection coefficient $Im(r)$, which describes the amplitudes of evanescent waves on the surface of the TI layer, for the entire system as a function of frequency $\omega$ and in-plane wave vector $k_{x}$. The imaginary part of the reflection coefficient is maximum whenever modes exist \cite{Duan2018,Woessner2015,Bezares2017,Kumar2015}, and thus this type of plot effectively reveals the dispersion relation. In the following sections we introduce the frequency-dependent formulas for the optical properties of the TI, the III-V, and the 2D interface state, all of which are required inputs to the scattering matrix. We then describe the scattering matrix approach employed. 

\subsection{DPP in a pristine Bi$_{2}$Se$_{3}$ thin film}
\label{DiracBiSe}

The layered van der Waals material Bi$_{2}$Se$_{3}$ with rhombohedral crystal structure is a topological insulator. Bi$_{2}$Se$_{3}$ has an insulating band gap as large as 0.3~eV in the bulk due to the strong spin-orbit coupling originating from the heavy element bismuth. The gapless surface states are characterized by a single Dirac cone in the electronic band structure with vanishing density of states near the Dirac point \cite{Moore2010,Hasan2010,Wang2020}, which enables the existence of two-dimensional massless Dirac fermions \cite{Pietro2013,Pietro2020}. These massless Dirac fermions are spin-momentum locked, which prohibits them from back-scattering into other surface states. The surface states of topological insulators, including Bi$_{2}$Se$_{3}$, can also host two dimensional spin-polarized DPPs. The behavior of these DPPs is analogous to that in graphene and the DPP system on the surface of pristine Bi$_{2}$Se$_{3}$ layer can be treated as a conducting electron sheet with optical conductivity given by \cite{Wang2020,Tse2010,Qi2014,Gusynin2006,Li2013}
\begin{align}
    \sigma_{{\rm Bi}_2{\rm Se}_3}&=\frac{e^{2}E_{F}}{4\pi \hbar^{2}}\frac{i}{\omega+i\tau^{-1}}.
\end{align}
The frequency-dependent permittivity of the bulk Bi$_{2}$Se$_{3}$ can be described by the Drude–Lorentz model, which gives \cite{Dordevic2013,Wang2020}:
\begin{widetext}
\begin{align}
    \varepsilon_{{\rm Bi}_2{\rm Se}_3}(\omega) = 1-\frac{\omega_{D}^{2}}{\omega^{2}+i\omega\gamma_{D}}+\frac{S_{\alpha}^{2}}{\omega_{\alpha}^{2}-\omega^{2}-i\omega\gamma_{\alpha}} + \frac{S_{\beta}^{2}}{\omega_{\beta}^{2}-\omega^{2}-i\omega\gamma_{\beta}}+ \frac{S_{G}^{2}}{\omega_{G}^{2}-\omega^{2}-i\omega\gamma_{G}}.
\end{align}
\end{widetext}
Here $E_{F} \approx 260~meV$ is the Fermi energy of surface states, $\tau \approx 0.06~ps$ is the relaxation time, $e$ is the electron charge. $\omega_{x},\gamma_{x}$, and $S_{x}$ represent the frequency, the scattering rate, and the strength of the Lorentz oscillator associated with the $\alpha$ phonon ($x=\alpha$), the $\beta$ phonon ($x=\beta$), the Drude-like bulk electrons ($x=D$), and the bulk band gap of the Bi$_{2}$Se$_{3}$ thin film ($x=G$). Numerical values for these parameters were acquired from fits to far and mid-IR transmittance measurements of the Bi$_{2}$Se$_{3}$ thin film at room temperature \cite{Wang2020}: $\omega_{D}=908.66~cm^{-1}$, $\gamma_{D}=7.43~cm^{-1}$, $\omega_{\alpha}=63.03~cm^{-1}$, $S_{\alpha}=675.9~cm^{-1}$, $\gamma_{\alpha}=17.5~cm^{-1}$, $\omega_{\beta}=126.94~cm^{-1}$, $S_{\beta}=100~cm^{-1}$, $\gamma_{\beta}=10~cm^{-1}$, $\omega_{G}=2029.5~cm^{-1}$, $S_{G}=11249~cm^{-1}$, and $\gamma_{G}=3920.5~cm^{-1}$ \cite{Wang2020}. Because the work reported here focuses on the spectral range between $1~THz$ and $9~THz$, we can neglect the contribution of the bulk gap, which is at very high frequency. Furthermore, we are imagining Bi$_{2}$Se$_{3}$ samples grown by molecular beam epitaxy (MBE), which results in low bulk carrier densities, allowing us to neglect the Drude term in equation \ref{permiBiSe} \cite{Wang2020}. As a result, only the contributions of the $\alpha$ and $\beta$ phonons play an important roles in the frequency-dependent expression of the Bi$_{2}$Se$_{3}$ permittivity, and our modeling uses the following reduced form:
\begin{align}
    \varepsilon_{{\rm Bi}_2{\rm Se}_3}(\omega) = 1+\frac{S_{\alpha}^{2}}{\omega_{\alpha}^{2}-\omega^{2}-i\omega\gamma_{\alpha}} + \frac{S_{\beta}^{2}}{\omega_{\beta}^{2}-\omega^{2}-i\omega\gamma_{\beta}}.
    \label{permiBiSe}
\end{align}

\subsection{III-V quantum well structure and intersubband transition}
\label{III_V_ISBT}

We now look at the effect of the intersubband transition (ISBT) in the AlGaAs/GaAs quantum well on the optical properties of the hybrid structure. We first need to determine the structural parameters that create a III-V quantum well (QW) with ISBTs in the spectral range between $1$ and $9~THz$. We consider a QW composed of an AlGaAs barrier and a GaAs well in which electrons can move freely parallel to the sample surface, creating a free two dimensional electron gas (2DEG) in the x-y plane. Quantum confinement along the growth (z) direction creates quantized energy states for electrons and the possibility for ISBTs between these quantized states in response to THz radiation with an electric field along the z direction. We use Al$_{0.4}$Ga$_{0.6}$As in order to provide 0.3~eV \cite{Wang2013,Vurgaftman2001} of confinement for electrons in the conduction band while staying within the regime in which Al$_x$Ga$_{1-x}$As has a conduction-band minimum at the $\Gamma$ point. We consider structures with a varying number of identical QWs. In the absence of coupling to the DPP, increasing the number of QWs in the structure increases the total absorption of energy resonant with the ISBTs. We analyze the dependence of the hybridized states on the number of QWs below. 

\begin{figure}[h]
\centering
    \includegraphics[width=.48\textwidth]{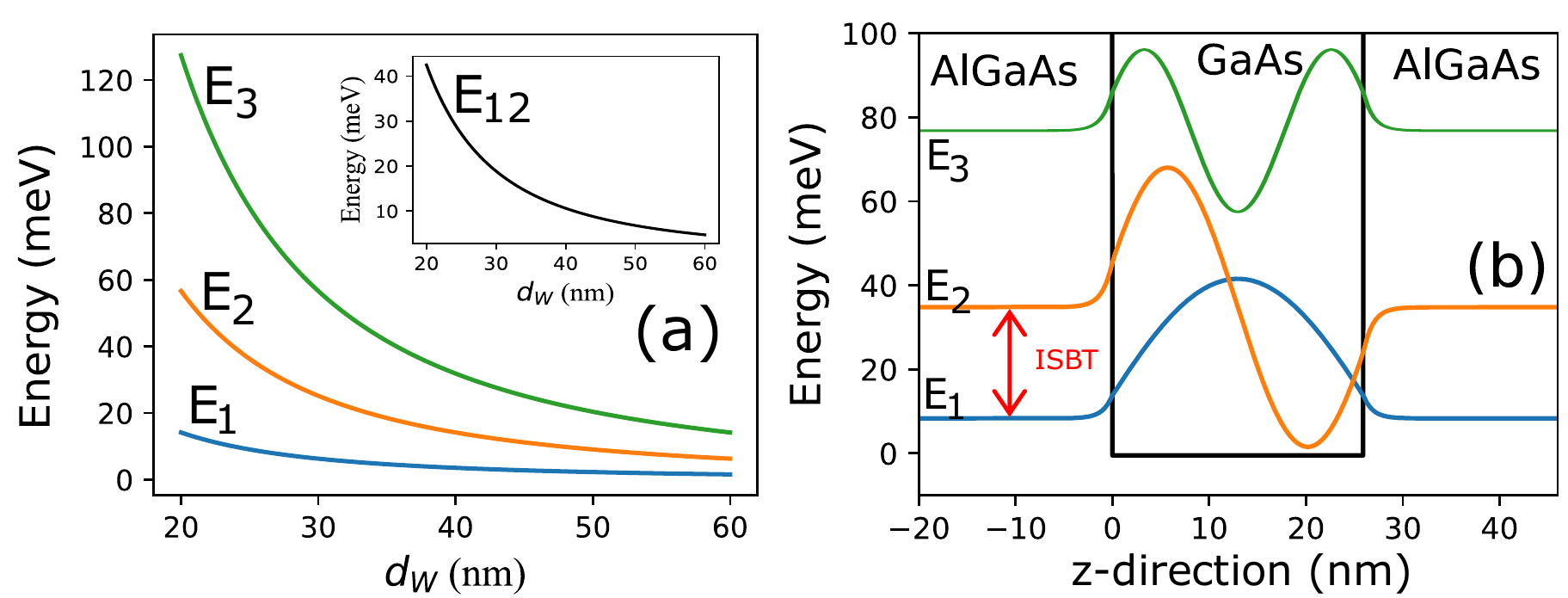}
 \caption{(a) Energies of the first three quantized levels of electrons in the conduction bands of the AlGaAs/GaAs quantum well as a function of the GaAs well thickness. The inset presents the well thickness-dependence of the transition energy between the first and second levels. (b) The spatial dependence of the wavefunction corresponding to the first three quantized energy states in the AlGaAs/GaAs quantum well solved using the effective mass approximation method.}
  \label{FIG2}
\end{figure}

Within the envelope function and effective mass approximations, the electronic states in the conduction band of a single III-V QW are well described by the Schr\"odinger equation:
\begin{equation}
    \left[ -\frac{\hbar^{2}}{2} \nabla \frac{1}{m^{*}(z)}\nabla + V(z) \right] \psi_{n} = E_{n}\psi_{n}(z)
\end{equation}
where $m^{*}(z)$ is the effective mass of the electrons in the conduction band and $V(z)$ is the potential landscape formed by the band offset between the AlGaAs and the GaAs materials. We use $m^{*}_{GaAs}=m^{*}_{AlGaAs}= 0.065$, $V=0$ for GaAs, and $V=0.3~eV$ for the Al$_{0.4}$Ga$_{0.6}$As barriers \cite{Wang2013,Vurgaftman2001}, which we take to be $100~nm$ thick. We choose thick Al$_{0.4}$Ga$_{0.6}$As barriers so that there is no coupling between adjacent QWs and no dependence of the ISBT energy on the number of QWs included in the structure. The Schr\"odinger equation can be solved exactly to provide the energies and wavefunctions of the quantized states as a function of $d_{W}$, the thickness of GaAs well \cite{Davies1998a, To2019}. The energies of the first three quantized levels (n=1,2,3) in this QW as a function of well thickness $d_{W}$ are shown in Fig.~\ref{FIG2}(a), and the wavefunctions associated with these three states for $d_{W}=25~nm$ are shown in Fig.~\ref{FIG2}(b). We see that the first and second states (E$_1$ and E$_2$) have opposite symmetry along the growth direction (z), which means that they will have a large optical dipole matrix element for interaction with radiation having an electric field along the z direction. The inset to Fig.~\ref{FIG2}(a) shows the energy difference between these two states as a function of $d_{W}$. A frequency of the ISBT between $1$ and $9~THz$ corresponds to an energy difference between $4$ and $36~meV$, which can be achieved for GaAs well thicknesses between $20$ and $60~nm$. In the rest of this work we primarily consider a frequency of $6~THz$ ($\approx 25~meV$) which corresponds to a well thickness of $25~nm$. 

We now construct the permittivity tensor for a III-V heterostructure containing multiple instances of this AlGaAs/GaAs QW. The asymmetry between the motion of the electrons along the z direction and in the x-y plane results in an anisotropic permittivity tensor. The frequency-dependence of the in-plane terms is established by using the Drude-Lorentz model, 
\begin{equation}
    \epsilon^{xx} = \epsilon^{yy} = \langle \epsilon^{xy} \rangle - \frac{N_{s}e^{2}}{\epsilon_{0}m^{*}L_{QW}}\frac{1}{\omega^{2}+i\omega\gamma_{1}},
\label{epsxx}
\end{equation}
and the z-component of the permittivity tensor involving the ISBTs is given by \cite{Zaluzny1998,Plumridge2007,Plumridge2008,Todorov2010,Shekhar2014}:
\begin{equation}
    \frac{1}{\epsilon^{zz}} =  \frac{1}{\left\langle\epsilon^{z}\right\rangle}  - \frac{\hbar^{2}}{\epsilon_{well}}\left[ \sum_{m,n} \frac{\omega_{P}^{2}f_{nm}}{E_{mn}^{2} -\hbar^{2} \left(\omega^{2}+ 2i\gamma_{2}\omega \right)}  \right].
    \label{epszz}
\end{equation}
In these expressions $\epsilon_{0}$ is the vacuum permittivity, $L_{QW}=d_{B}+d_{W}$ is the thickness of a single AlGaAs/GaAs QW, $N_{s}$ is the sheet carrier concentration originating in the doping level in the AlGaAs/GaAs quantum well, and $\gamma_{1}=\gamma_{2}$ are the in- and out-of-plane electron-scattering rates of the ISBT; $\omega_{P}$ is the plasma frequency, given by:
\begin{equation}
    \omega_{P} = \sqrt{\frac{ N_{s}e^{2}}{\epsilon_{0}m^{*}L_{QW}}}.
\end{equation}
$\langle \epsilon^{xy} \rangle$ and $\langle \epsilon^{z} \rangle$ represent the mean effective background dielectric constants for, respectively, the in-plane and out-of-plane permittivity, and are given by:
\begin{equation}
    \langle \epsilon^{xy} \rangle = \left(1 - \frac{L_{QW}}{L_{MQW}} \right) \epsilon_{B}
\end{equation}
and 
\begin{equation}
     \frac{1}{\left\langle\epsilon^{z}\right\rangle}  = \frac{1}{\epsilon_{B}} \left( 1- \frac{L_{QW}}{L_{MQW}} \right) + \frac{1}{\epsilon_{well}} \frac{L_{QW}}{L_{MQW}},
\end{equation}
where $\varepsilon_{B}$ and $\varepsilon_{Well}$ are the dielectric constants for the undoped barrier and well compositions, respectively, and $L_{MQW}$ is total thickness of the multiple QW part of the structure ($L_{QW}$ times the total number of quantum wells). The sum in equation \ref{epszz} is taken over all allowed transitions and the oscillator strength of the each transition, $f_{mn}$, is given by:
\begin{equation}
    f_{mn} = \frac{2m^{*}E_{mn}}{\hbar^{2}}z_{mn}^{2},
\end{equation}
with $z_{mn} = \langle m \vert z \vert n \rangle$ the intersubband dipole matrix element that can be estimated by \cite{Helm19991}:
\begin{equation}
    f_{mn} = \frac{64\left(mn\right)^{2}}{\pi^{2}\left(n^{2}-m^{2}\right)^{3}},
\end{equation}
resulting in $f_{12} \approx 1$ for a transition taking place between the first and second quantized levels of this QW design. We note that the scattering rate parameters $\gamma_{1}$ and $\gamma_{2}$ depend heavily on crystalline and interface quality, which are specific to individual samples. These parameters should thus be determined experimentally. In the first part of this work we use $\gamma_{1} \sim \gamma_{2} \sim 1~THz$, which ensures that the loss rates are not too large in comparison to the interaction between the resonators, allowing us to explore the coupling between the excitations in the $THz$ frequency regime. In a later section we consider how variations in $\gamma_{1}$ and $\gamma_{2}$ will alter the results.

\subsection{Two dimensional hole gas at the TI/III-V interface from DFT}
\label{DFTpredict}

The interactions between the DPP at the TI top surface and the ISBTs in the QWs can be affected by the interface states at the Bi$_{2}$Se$_{3}$/GaAs(001) interface. Since no experimental information exists for the electronic structure of this interface, we carried out density functional theory (DFT) calculations taking into account the possible interface terminations and the resulting interface electronic structure.
Our calculations are based on DFT \cite{Kohn1964,Kohn1965} within the generalized gradient approximation (GGA) PBEsol \cite{Perdew2008,Perdew2009}, and projector augmented wave (PAW) potentials \cite{Bloch1994} as implemented in the Vienna Ab initio Simulation Package (VASP) \cite{Kresse1996,Kresse1999}. Van der Waals interactions are included using the D3 method of Grimme et al., \cite{Grimme2010}. 

\begin{figure}[h]
\centering
    \includegraphics[width=.5\textwidth]{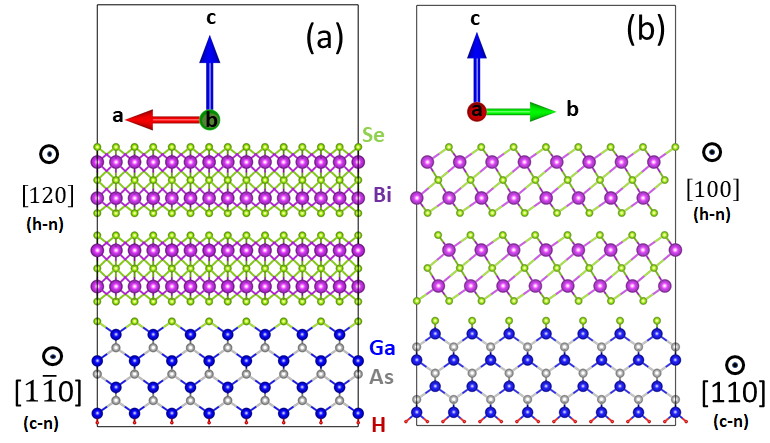}
 \caption{Atomic structures of the Bi$_2$Se$_3$/GaAs interface along (a) c-axis, (b) a-axis together with the indications of hexagonal notation (h-n) for Bi$_2$Se$_3$ and cubic notation (c-n) for GaAs.}
  \label{FIG13}
\end{figure}

We use a plane wave cutoff energy of $400~eV$ for the plave wave expansions, and a $\Gamma$-centered mesh of $7 \times 1 \times 1$ special $k$ points for the integrations over the Brillouin zone for a supercell representing the Bi$_{2}$Se$_{3}$/GaAs(001) interface in a slab geometry. This interface supercell contains two quintuple layers of Bi$_{2}$Se$_{3}$ and four layers of GaAs bilayers with the Ga-terminated bottom layer passivated by pseudo-hydrogen atoms with charge of $1.25~e$. To accommodate the lattice mismatch between Bi$_{2}$Se$_{3}$ and GaAs(001), we strained the Bi$_{2}$Se$_{3}$ layer by $1.0~\%$ and $3.4~\%$ in the$[110]$ and $[1\overline{1}0]$ direction, respectively (the hexagonal notation for the Bi$_2$Se$_3$ is indicated in Fig. \ref{FIG13}). In this supercell, there are 7 As atoms on the top surface of the GaAs(001) slab in one supercell. A vacuum spacing of over $15\AA$ was used to minimize the spurious interaction between the slab and its image along the direction perpendicular to the interface. All atoms, except the bottom Ga-H layers, were allowed to relax until the residual force on each atom was less than 0.01 $eV/\AA$. 

\begin{figure}[h]
\centering
    \includegraphics[width=.4\textwidth]{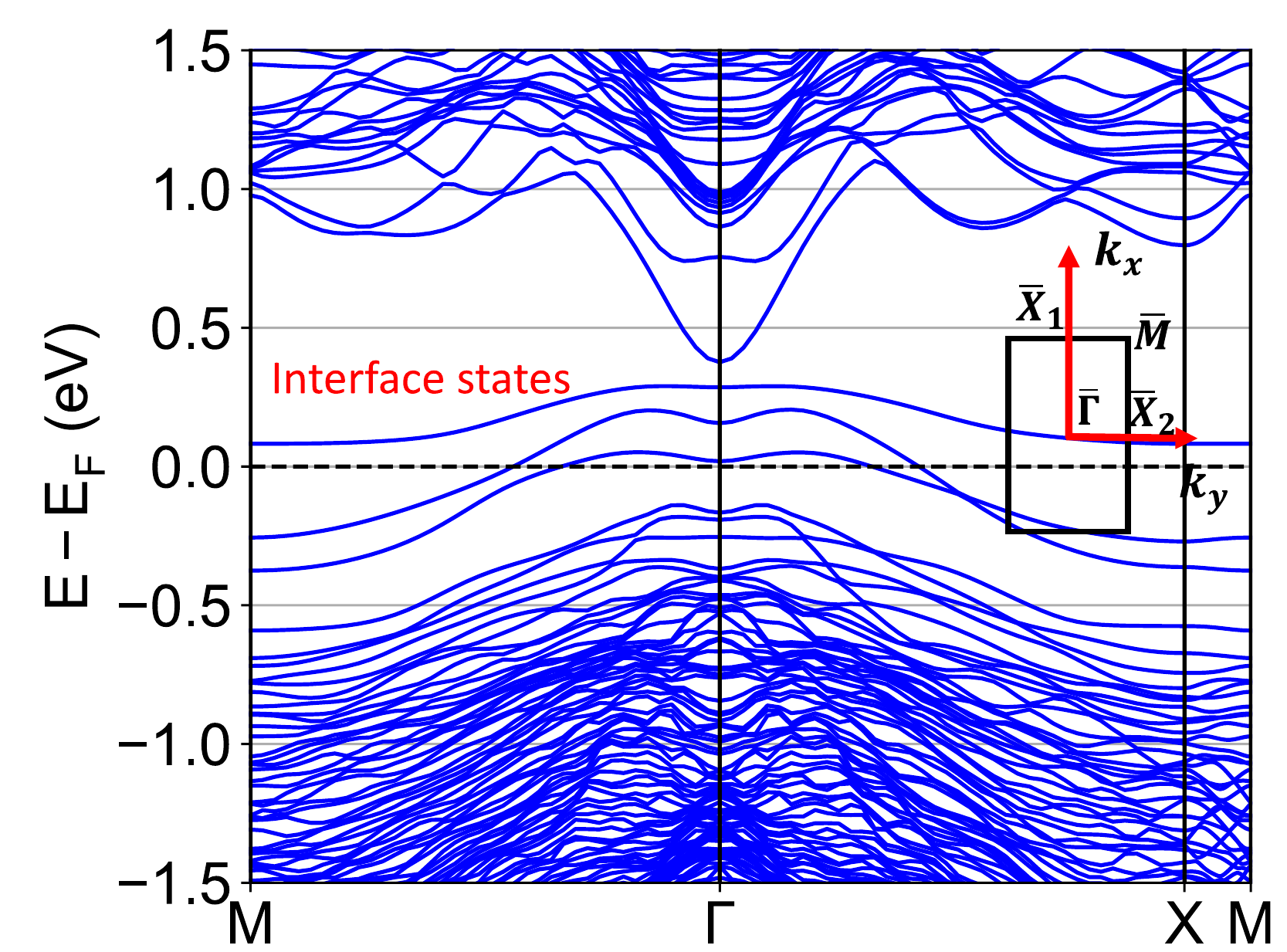}
 \caption{Calculated electronic band structure of the Bi$_{2}$Se$_{3}$/GaAs interface. The Fermi level is set to zero in the energy axes. The inset represents the first Brillouin zone of the Bi$_{2}$Se$_{3}$/GaAs interface.}
  \label{FIG3}
\end{figure}
We explored three different terminations for the Bi$_{2}$Se$_{3}$/GaAs(001) interface, namely Ga-, As-, and Se-terminated GaAs(001). Our calculated interface formation energies as a function of As and Se chemical potentials at two limiting conditions (As-rich/Se-poor and As-poor/Se-rich) reveal lower formation energies for the interfaces terminated with Se. This is due to the fact that the As dangling bonds are partially occupied with 5/4 electrons each, and thus tend to form As-As dimers \cite{Liu2022}. By replacing the top As monolayer with Se, each Se dangling bond will carry 7/4 electrons. This almost-completely-filled dangling bond configuration precludes the formation of Se-Se dimers in the Se-terminated Bi$_{2}$Se$_{3}$/GaAs(001) interface, which is in good agreement with the TEM images obtained for such an interface and consistent with the growth conditions in which selenization of the GaAs(001) interface was performed before deposition of  Bi$_{2}$Se$_{3}$ \cite{Liu2022}. Therefore, we conclude that the interface is Se terminated. 

The atomic structure of the Bi$_{2}$Se$_{3}$/GaAs(001) interface is shown in Fig.~\ref{FIG13} and the electronic structure of the Bi$_{2}$Se$_{3}$/GaAs(001) interface with Se terminated GaAs(001) is shown in Fig.~\ref{FIG3}. The Fermi level, $E_F$, is set at zero on the energy axis.  We observe interface states derived mostly from Se (4$p$ orbitals) at the interface layer. These interface states cross the Fermi level, indicating the formation of a 2DHG with high carrier density.  The charge density plots of these interface states indicate that the 2DHG is tightly bound to the interface, localized to less than 5 atomic layers, and highly delocalized in the plane of the interface.
The existence of this 2DHG at the Bi$_{2}$Se$_{3}$/GaAs(001) interface adds an additional term in the Drude model for the optical conductivity of the Bi$_{2}$Se$_{3}$ surface plasmon, i.e.,
\begin{equation}
    \sigma_{2DHG} = \frac{e^{2}N_{2DHG}}{m^{*}_{2DHG}\left( \frac{1}{\tau_{2DHG}}-i\omega \right)},
    \label{extcond}
\end{equation}
where $N_{2DHG}$ is the carrier concentration in the 2DHG, $m^{*}_{2DHG}$ is the effective mass of holes around the Fermi energy at the Bi$_{2}$Se$_{3}$/GaAs(001) interface, and $\tau_{2DHG}$ is the scattering time. Noting the relatively small dispersion of the interface states, we expect the mass of the holes to be significantly larger than that in GaAs. Here we approximate $m^{*}_{2DHG} \approx 1$, i.e., the same as the rest mass of the electron. Deviations from this approximation are discussed in later sections.

\subsection{Scattering matrix method for studying hybridized excitations of the TI and III-V}
\label{ScatteringMethod}

Now that we have constructed the optical response functions for the hybrid material schematically shown in Fig.~\ref{FIG1}), we investigate the coupling between the TI and the III-V constituents by solving Maxwell's equations to derive the dispersion relationship for a monochromatic electromagnetic (EM) wave propagating in our optical structure. We do this using the scattering matrix formalism that has proven to be a powerful tool for investigating the optical properties of layered structures \cite{Katsidis2002}. We present a detailed description of the scattering matrix formalism in Appendix~\ref{Smatrix}. The most important outcome of this formalism for the work presented here is that we can compute the optical response of the entire structure from a scattering matrix constructed using a recursive method based on interface and propagation matrices that capture what happens at each interface and within each layer of the structure \cite{To2019}. The inputs to these interface and propagation matrices are the materials parameters of the system and the permittivity within each layer. 

From the permittivities derived in the previous sections, we employ the scattering matrix formalism to compute the reflection coefficients for our hybrid material system. The imaginary part of the reflection coefficient, $Im(r)$, is proportional to the losses in the system. The presence of loss in the reflectance spectrum indicates that the incident EM wave has generated an excitation that is carrying energy away laterally, i.~e., propagating in the x- or y-direction rather than transmitting or reflecting in the +z or -z directions, respectively. The frequency dependence of such loss thus generates the dispersion curves for the hybridized excitations in the coupled system, which is the aim of this study. In the next section we consider how this dispersion relation depends on structural and material properties, which allows us to probe the physics underlying the formation of hybridized excitations.

\section{Results and Discussion}
\label{results}

The scattering matrix technique is applied to study the structure shown in Fig.~\ref{FIG1}. In this model, light with both s- and p-polarized components is incident on the top Bi$_{2}$Se$_{3}$ layer. As a consequence of the electrostatic interaction with the electric field of the light, surface Dirac plasmon polaritons (DPPs) will be excited in the Bi$_{2}$Se$_{3}$ layer at certain resonant frequencies. These DPPs can then couple with the phonon in the bulk of the Bi$_{2}$Se$_{3}$ and also couple to the ISBTs in the AlGaAs/GaAs QW. TE-polarized light cannot excite the DPPs because it does not satisfy the boundary conditions. Consequently we consider only TM-polarized incident light in the our analysis. Throughout our analysis the structural parameters of a single AlGaAs/GaAs QW are kept fixed; the AlGaAs barrier thickness $d_{B}=100~nm$ and GaAs well thickness $d_{W}=25~nm$ give rise to a $6~THz$ frequency for the ISBT between the first and second quantized levels of the QW. The color plots in the following figures show the imaginary part of the Fresnel reflection coefficient $Im(r)$ of the entire structure; the maxima of the function $Im(r)$ reveals the dispersion relationship for the coupled modes. We first discuss the emergence and characteristics of coupled modes and then consider how the strength of the coupling depends on structural and material parameters. We note that Table \ref{table1} in Appendix \ref{parameters} contains a complete list of the computational parameters used in all figures we present here.  

\subsection{Signatures of coupling between DPP and ISBT}
\begin{figure}[htb]
\centering
    \includegraphics[width=.5\textwidth]{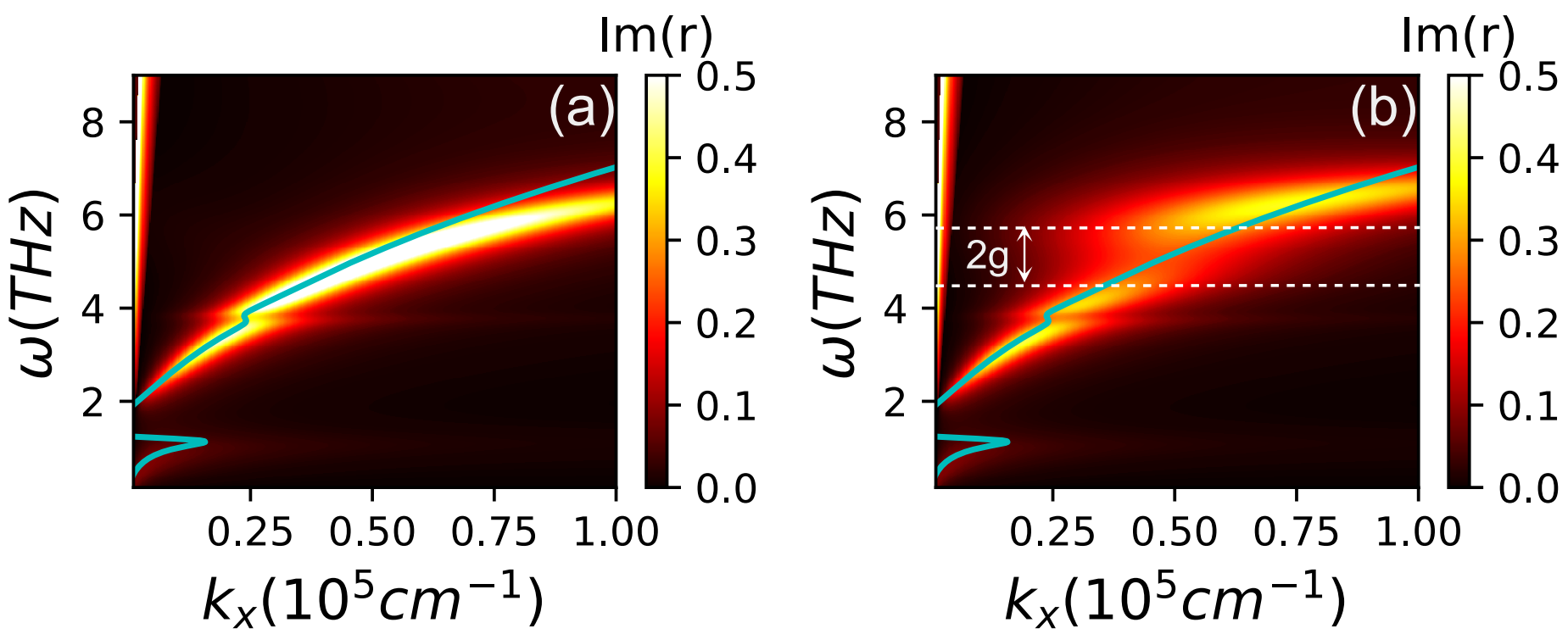} 
 \caption{The dispersion relation of the Dirac plasmon-phonon-ISBT polarition for the structure depicted in Fig.~\ref{FIG1} with a single AlGaAs/GaAs quantum well as a function of the frequency $\omega$ and the in-plane wavevector $k_{x}$. The dispersion relation is calculated for two different thicknesses of the GaAs spacer layer, (a) $d_{sp}=500~nm$ and (b) $d_{sp}=10~nm$. The Bi$_{2}$Se$_{3}$ layer has thickness $d_{TI}=100~nm$. The parameters of the AlGaAs/GaAs QW are $d_{B}=100~nm$ and $d_{w}=25~nm$, which give rise to an ISBT at $6~THz$, and a doping level of $N_{s}=10^{12}cm^{-2}$. The solid blue line provides, for reference, an analytical calculation of the dispersion of the DPP in a pristine Bi$_{2}$Se$_{3}$ layer on a GaAs substrate, as described in the text. The inset to (b) indicates how the coupling strength, g, is extracted from the calculated anticrossing magnitude.}
  \label{FIG5}
\end{figure}

We start by considering the interaction between the TI layer and a single III-V QW structure via the dispersion relation. The color plot in Fig.~\ref{FIG5} displays the imaginary part of the Fresnel reflection coefficient $Im(r)$ calculated for the entire structure as a function of the the frequency $\omega$ and the in-plane wavevector $k_{x}$. The dispersion relation is obtained by applying the scattering matrix method described in Sect.~\ref{ScatteringMethod} to a structure with Bi$_{2}$Se$_{3}$ layer thickness $d_{TI}=100~nm$ for two different thicknesses of the GaAs spacer layer: (a) $d_{sp}=500~nm$ and (b) $d_{sp}=10~nm$. The solid blue curve is an analytical calculation of the dispersion of the DPP in a pristine Bi$_{2}$Se$_{3}$ layer on a GaAs substrate, given by:
\begin{equation}
    \omega_{{\rm Bi}_{2}{\rm Se}_{3}}^{2} = \frac{e^{2}v_{F}\sqrt{2\pi n_{2D}}}{\varepsilon_{0}h} \left(\frac{k_{||}}{\varepsilon_{1} + \varepsilon_{3}+ k_{||}d_{TI}\varepsilon_{2}} \right),
    \label{BiSedispersion}
\end{equation}
where $v_{F} = 5 \times 10^{5}~m/s$ is the Fermi velocity for the DPP in Bi$_{2}$Se$_{3}$; $n_{2D}=1.2 \times 10^{13}~cm^{-2}$ is the sheet carrier concentration of the entire Bi$_{2}$Se$_{3}$ thin film, including the contribution from both surfaces \cite{Wang2018,Wang2020}; $\varepsilon_{1}$, $\varepsilon_{2}$ and $\varepsilon_{3}$ are the permittivity of the top, middle, and bottom dielectric media, respectively, corresponding to air, bulk Bi$_{2}$Se$_{3}$, and the GaAs substrate, respectively; $k_{||}$ is the in-plane wavevector; and $d_{TI}$ is the thickness of the Bi$_{2}$Se$_{3}$ layer. 

In Fig.~\ref{FIG5}(a) we plot the dispersion relation for a sample with a GaAs spacer layer of $500~nm$, which is sufficiently thick that the ISBT in the QWs has negligible coupling to the excitations in the Bi$_{2}$Se$_{3}$ layer. As a result, we observe only the dispersion of the Dirac plasmon-phonon polaritons of a single layer of Bi$_{2}$Se$_{3}$ on a GaAs substrate. Predictably, this DPP dispersion is comparable to the analytic result reported by Wang {\em et al.}, who investigated the plasmon dispersion in a pristine Bi$_{2}$Se$_{3}$ thin film on a sapphire substrate \cite{Wang2020}. In other words, Fig.~\ref{FIG5}(a) simply verifies that the scatting matrix approach (color plot) agrees with the analytical dispersion (solid line) when applied to a sample in which interactions with the QW are suppressed by a thick spacer layer. Both the analytical and scattering matrix approach predict the observed signatures of the coupling between the DPP on the surface of Bi$_{2}$Se$_{3}$ with the $\alpha$ and $\beta$ phonons in the bulk Bi$_{2}$Se$_{3}$ at frequencies around $1.9~ THz$ and $4~THz$ \cite{Wang2020,Pietro2013}.

In Fig.~\ref{FIG5}(b) we reduce the thickness of the GaAs spacer to $d_{sp}=10~nm$. We observe a significant change in the dispersion relation around $5~THz$ due to the interactions of the DPP with the ISBT in the single AlGaAs/GaAs QW. The DPP-ISBT coupling results in a splitting, which is evident through the reduction of the value of $Im(r)$ at approximately $k_{x}=0.5 \times 10^5$ $cm^{-1}$ and $\omega=5~THz$ and the increased $Im(r)$ at approximately $\omega=6~THz$ for $k_{x}<0.5 \times 10^5$ $cm^{-1}$. The magnitude of the splitting that arises due to the coupling between the DPP and the ISBT is evaluated by looking at the separation between the upper and lower polariton branches as indicated by the separation between the two dashed white lines in Fig.~\ref{FIG5}(b). The splitting $(2g)$ is twice the strength of the coupling, giving $g \approx 0.7~THz$. This splitting should be experimentally detectable because it is comparable to the linewidth of the isolated modes in the system. The full-width-half-max (FWHM) linewidths of the $\alpha$ and $\beta$ phonons in the TI layer are both of order $1~THz$ and the FWHM linewidth of the plasmon on the surface of the TI is around $3~THz$. The FWHM of the DPP shown in Fig.~\ref{FIG5}(a) is $0.88~THz$. The FWHM linewidth of the ISBT in the QW is also of order $1~THz$. Because $g$ is comparable to the linewidth of both excitations that are being hybridized (DPP and ISBT), $g \approx 0.7~THz$ indicates the formation of a hybridized state that is approaching the strong coupling regime. 

In the presence of the splitting due to DPP-ISBT interactions there are four separate polariton branches in the dispersion relation. For the convenience of discussion in the rest of this paper, we will refer to the modes above the ISBT ($\omega>5~THz$) as the upper polariton branch, the modes between the $\beta$ phonon and ISBT ($4<\omega<5~THz$) as the first middle polariton branch, the modes between the $\alpha$ and $\beta$ phonons ($2<\omega<4~THz$) as the second middle polariton branch, and the modes below the $\alpha$ phonon ($\omega<2~THz$) as the lower polariton branch.

\subsection{Role of TI phonons}
\label{phonons}
\begin{figure}[htb]
\centering
    \includegraphics[width=.5\textwidth]{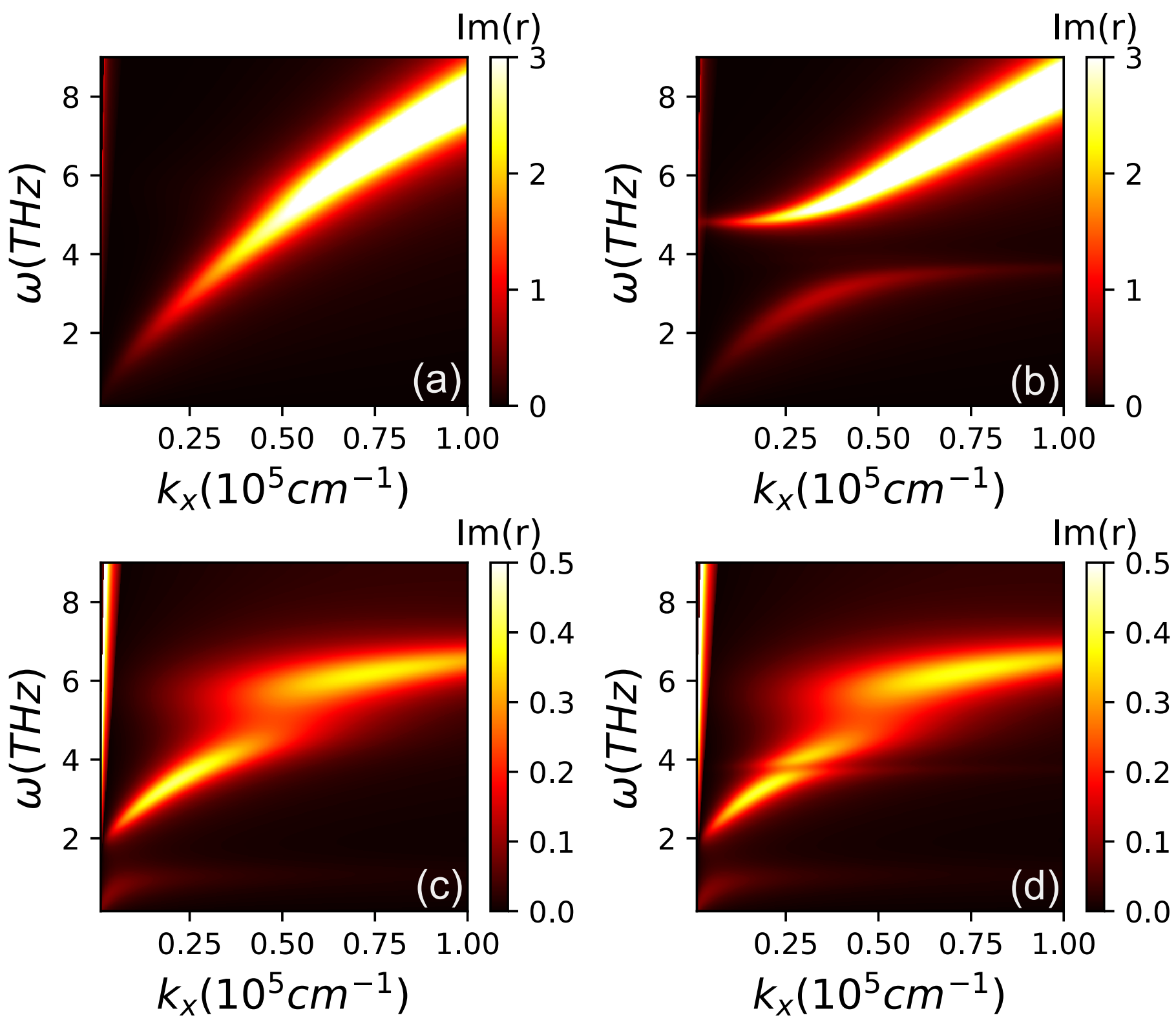} 
 \caption{The dispersion relation calculated (a) without the phonons in the bulk TI, (b) with only the $\beta$ phonon at $\sim4~THz$, (C) with only the $\alpha$ phonon at $\sim2~THz$, and (d) with both the $\alpha$ and the $\beta$ phonons. The structural parameters of the heterostructure are $d_{TI}=100~nm$, $d_{B}=100~nm$,$d_{w}=25~nm$, $d_{sp}=10~nm$, and $N_{s}=10^{12}cm^{-2}$.}
  \label{PhononFig}
\end{figure}

The dispersion relations shown in Fig.~\ref{FIG5} involve a plasmon-phonon-polariton in the bare TI without (Fig.~\ref{FIG5}(a)) and with (Fig.~\ref{FIG5}(b)) additional hybridization with the ISBT of the QW. Next, we consider the importance of the phonons. In Fig.~\ref{PhononFig} we compute the dispersion relation by mathematically turning on and off the contributions of the $\alpha$ and the $\beta$ phonons in the bulk of the TI. All calculations are done for the conditions otherwise identical to Fig.~\ref{FIG5}(b), meaning that the GaAs spacer layer is sufficiently thin ($d_{sp}=10~nm$) to result in a splitting due to strong coupling to the ISBT in the QW. Interestingly, we found that the $\alpha$ phonon plays a critical role in the formation of an observable splitting. In Fig.~\ref{PhononFig}(a) and (b) the $\alpha$ phonon at $\sim2~THz$ is turned off and the $\beta$ phonon at $\sim4~THz$ is (a) turned off and (b) turned on. A clear splitting due to coupling between the DPP and the $\beta$ phonon is evident in Fig.~\ref{PhononFig}(b), but in neither case is there evidence of a splitting in the vicinity of $\omega=6~THz$, which is where the signature of interaction with the ISBT would appear. In Fig.~\ref{PhononFig}(c) and (d) the $\alpha$ phonon at $\sim2~THz$ is turned on and the results are plotted with the $\beta$ phonon at $\sim4~THz$ (c) turned off and (d) turned on. In both cases we observe the splitting in the vicinity of $\omega=6~THz$ due to strong coupling to the ISBT in the QW. 

\begin{figure}[htb]
\centering
    \includegraphics[width=.38\textwidth]{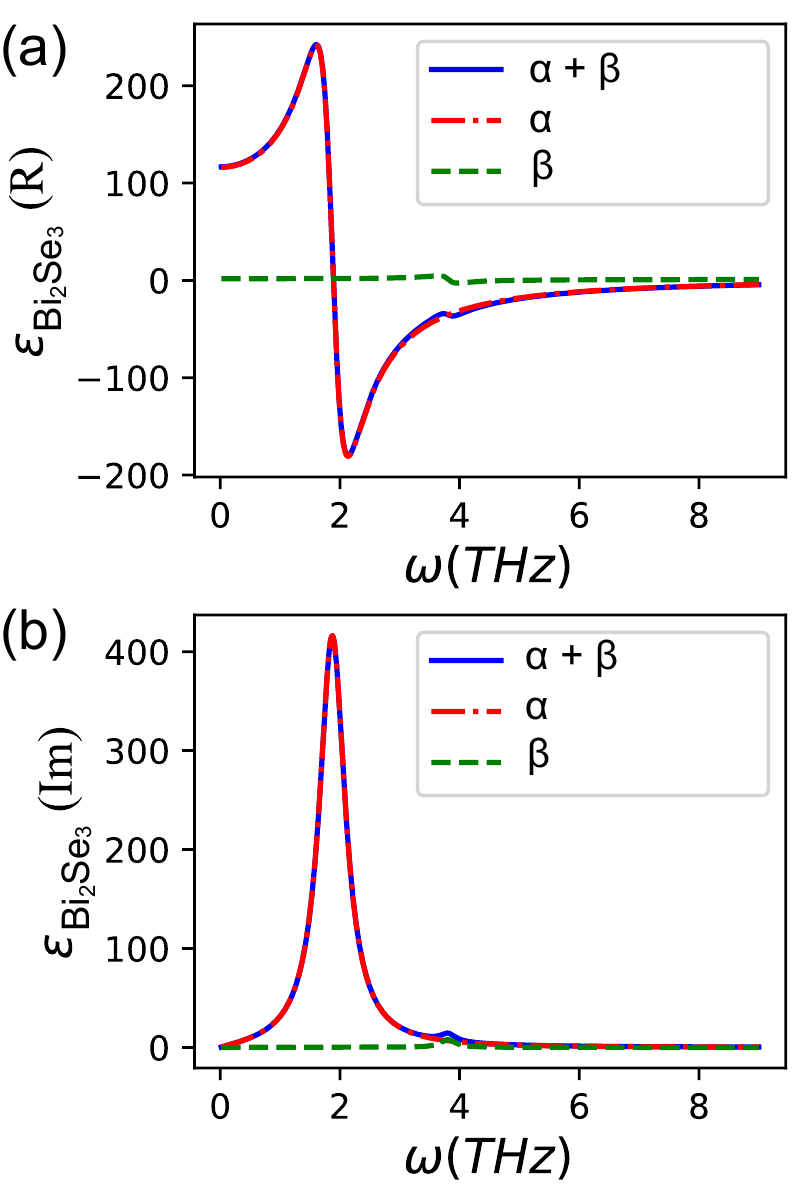} 
 \caption{The frequency dependent permittivity of Bi$_{2}$Se$_{3}$ with (a) the real (R) part and (b) the imaginary (Im) part computed by using equation \ref{permiBiSe} with the contributions from both alpha and beta phonons (blue); alpha phonon (red) and beta phonon (green) respectively indicated in the legends.}
  \label{Permittivity}
\end{figure}

Note that Fig.~\ref{FIG5}(a) provides a reference for the dispersion relation with both phonons turned on, but with the interaction with the ISBT suppressed by a thick GaAs spacer. These results predict that there will be no observable coupling between the DPP in the TI and the ISBT in the III-V QW in the absence of the $\alpha$ phonon resonance in the bulk TI layer. This occurs because the $\alpha$ phonon dominates the permittivity of Bi$_2$Se$_3$. In the absence of the $\alpha$ phonon the Bi$_2$Se$_3$ permittivity shifts in such a way that the DPP and ISBT are no longer in resonance. In contrast, the contribution of the $\beta$ phonon to the entire system is negligible in comparison to the $\alpha$ phonon. In Fig.\ref{Permittivity}(a) [b] we show the real [imaginary] part of the permittivity of the Bi$_2$Se$_3$ layer as a function of the frequency $\omega$ of an electromagnetic wave computed using Eq.~\ref{permiBiSe}. One observes that when the $\alpha$ phonon is suppressed, both the real part and imaginary part of the permittivity of the bare Bi$_2$Se$_3$ layer drop off dramatically, showing the minor contribution of the $\beta$ phonon. We note that it would be impossible to suppress the $\alpha$ phonon resonance experimentally.

\subsection{Dependence of the coupling strength on structure parameters}

\begin{figure}[htb]
\centering
    \includegraphics[width=.5\textwidth]{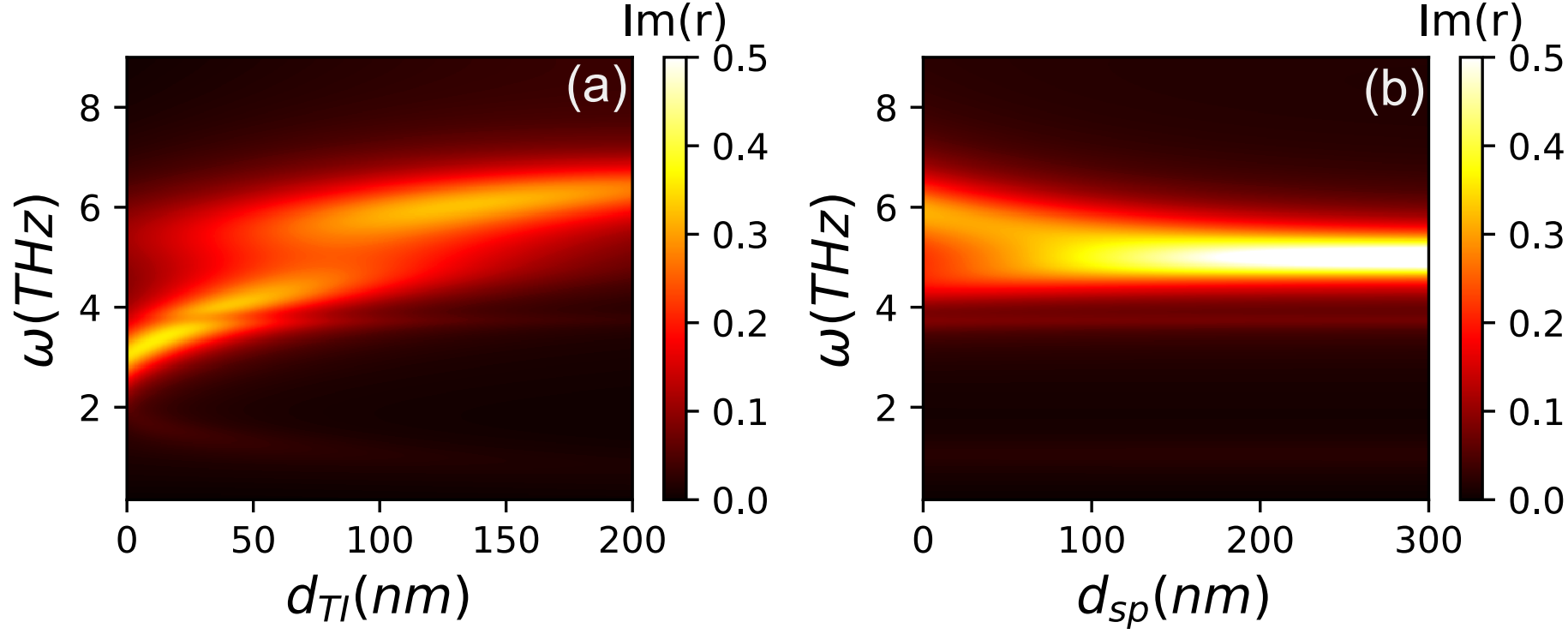}
 \caption{Dependence of the Dirac plasmon-phonon-ISBT polariton (DPP) modes on the thickness of (a) the Bi$_{2}$Se$_{3}$ and (b) the GaAs spacer layer. The calculations are performed for a specific wavevector $k_{x}=0.5 \times 10^{5} ~ cm^{-1}$ and with $N_{s}=10^{12}cm^{-2}$. In (a) the thickness of the GaAs spacer layer is kept fixed at $d_{sp}=10~nm$; in (b) the thickness of the TI layer is kept fixed at $d_{TI}=100~nm$.}
  \label{FIG6}
\end{figure}

We now consider the effect of changing the structure parameters. In Fig.~\ref{FIG6} we show the dependence of the DPP-ISBT polariton modes on the thickness of (a) the Bi$_{2}$Se$_{3}$ and (b) the GaAs spacer layer. The calculations were performed for the specific in-plane wavevector $k_{x}=0.5 \times 10^{5}~cm^{-1}$. In Fig.~\ref{FIG6}(a) we observe that the plasmon modes blueshift upon increasing the TI thickness, exactly the trend expected from the DPP in the bare Bi$_{2}$Se$_{3}$ film as described in Eq.~\ref{BiSedispersion}. This shift is caused by the dependence of the negative real part of the Bi$_{2}$Se$_{3}$ permittivity on the film thickness. As expected, we observe a splitting at a Bi$_{2}$Se$_{3}$ film thickness of $100~nm$, which is the film thickness used in the calculations reported and discussed above. In Fig.~\ref{FIG6}(b) we observe the effect of the GaAs spacer layer thickness. For GaAs spacer layer thicknesses larger than $\sim100~nm$ we observe a faint mode at $3.9~THz$ and a bright mode at $4.5~THz$. These are the modes of the phonon-plasmon polariton in the Bi$_{2}$Se$_{3}$ film and they do not change because the interactions with the ISBT in the QW are suppressed due to the thick GaAs spacer. As the GaAs spacer layer becomes thinner than $\sim100~nm$, the modes begin to shift. This result is expected: as the GaAs spacer layer becomes thinner the strength of the coupling between the plasmon-phonon polariton and the ISBT increases, leading to an increasingly large mode shift due to an increasing splitting. The most important result of these calculations is that the thinnest possible GaAs spacer layer thickness will lead to the largest splitting and one should not expect to observe coupling for a GaAs spacer layer thickness exceeding $\sim100~nm$.

\subsection{Dependence on material quality and scattering rates}

\begin{figure}[htb]
\centering
    \includegraphics[width=.5\textwidth]{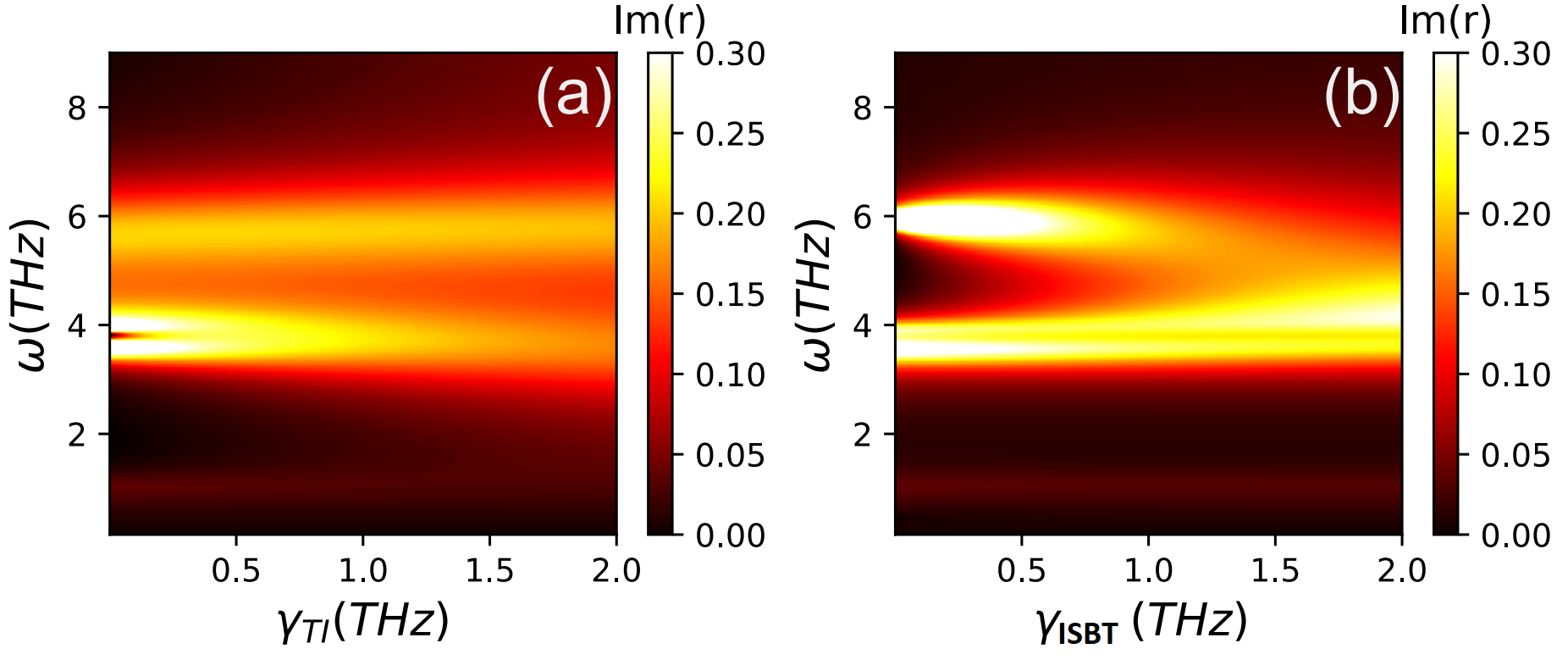}
 \caption{Dependence of the Dirac plasmon-phonon-ISBT polariton modes on (a) the scattering rate in the Bi$_{2}$Se$_{3}$ layer and (b) the scattering rate in the AlGaAs/GaAs QW. The calculations are performed for a specific wavevector $k_{x}=0.3 \times 10^{5} ~ cm^{-1}$, with $N_{s}=1.5 \times 10^{12}cm^{-2}$, and with the thicknesses of the Bi$_{2}$Se$_{3}$ and the GaAs spacer respectively $d_{TI}=100~nm$ and $d_{sp}=10~nm$.}
  \label{FIG7}
\end{figure}

The coupling between light and matter can only be observed when the coupling is large enough to be comparable to the energy loss in the system. For this reason we now investigate the consequence of the energy loss, which is characterized by the scattering rate (equivalently, linewidth) of the modes that interact to create the hybridized polariton. These scattering rates are largely determined by the structural quality of the material regarding defect and interface scattering or defect-mediated nonradiative recombination. By computationally changing the scattering loss rates in the TI and the III-V QW we can therefore understand how the quality of the TI and the III-V materials impact the splitting and determine the linewidths required to observe strong coupling in this system. For simplicity, in this analysis, we assume that all resonances in the TI material can be described by the same scattering rate; we make the same assumption for the III-V QW. In other words, when considering the impact of the TI scattering rate we replace both $\gamma_{\alpha}$ and $\gamma_{\beta}$ in Eq.~\ref{permiBiSe} with $\gamma_{TI}$, keep all $\gamma$s for the QW system fixed at $1~THz$, and plot the resulting modes as a function of $\gamma_{TI}$. The results are shown in Fig.~\ref{FIG7}(a). Similarly, when considering the impact of the scattering rate in the QW, we replace both $\gamma_1$ and $\gamma_2$ in Eq.~\ref{epsxx} and \ref{epszz} with $\gamma_{ISBT}$, keep $\gamma_{\alpha}$ and $\gamma_{\beta}$ fixed at the values provided in Sect.\ref{DiracBiSe}, and plot the resulting modes as a function of $\gamma_{ISBT}$. The results are shown in Fig.~\ref{FIG7}(b).

In Fig.~\ref{FIG7}(a) we observe that the scattering rate in the TI layer has almost no impact on the upper polariton mode at $\omega=6~THz$, which is the signature of the strong coupling between the Dirac plasmon-phonon polariton and the ISBT, mediated by the $\alpha$ phonon as discussed in Sect.~\ref{phonons}. We do observe an additional splitting in the polariton mode around $\omega=4~THz$, which is in the vicinity of the $\beta$ phonon, when $\gamma_{TI}<\sim0.2~THz$. This is a further indication that the interactions with the $\beta$ phonon are quite weak and implies that new spectral features might emerge only for extremely high TI material quality (small linewidth). In Fig.~\ref{FIG7}(b) we observe that the upper polariton mode disappears when $\gamma_{ISBT}$ exceeds $1~THz$. This occurs when the loss in the QW exceeds the coupling strength and provides a benchmark for the QW scattering rate that would need to be achieved to create an experimentally observable coupling between the TI and the III-V QW. Taken as a whole, this analysis shows the critical importance of high quality samples for an experimental investigation of the strong coupling regime explored theoretically here.

\subsection{Dependence on QW doping density}

 \begin{figure}[htb]
\centering
    \includegraphics[width=.35\textwidth]{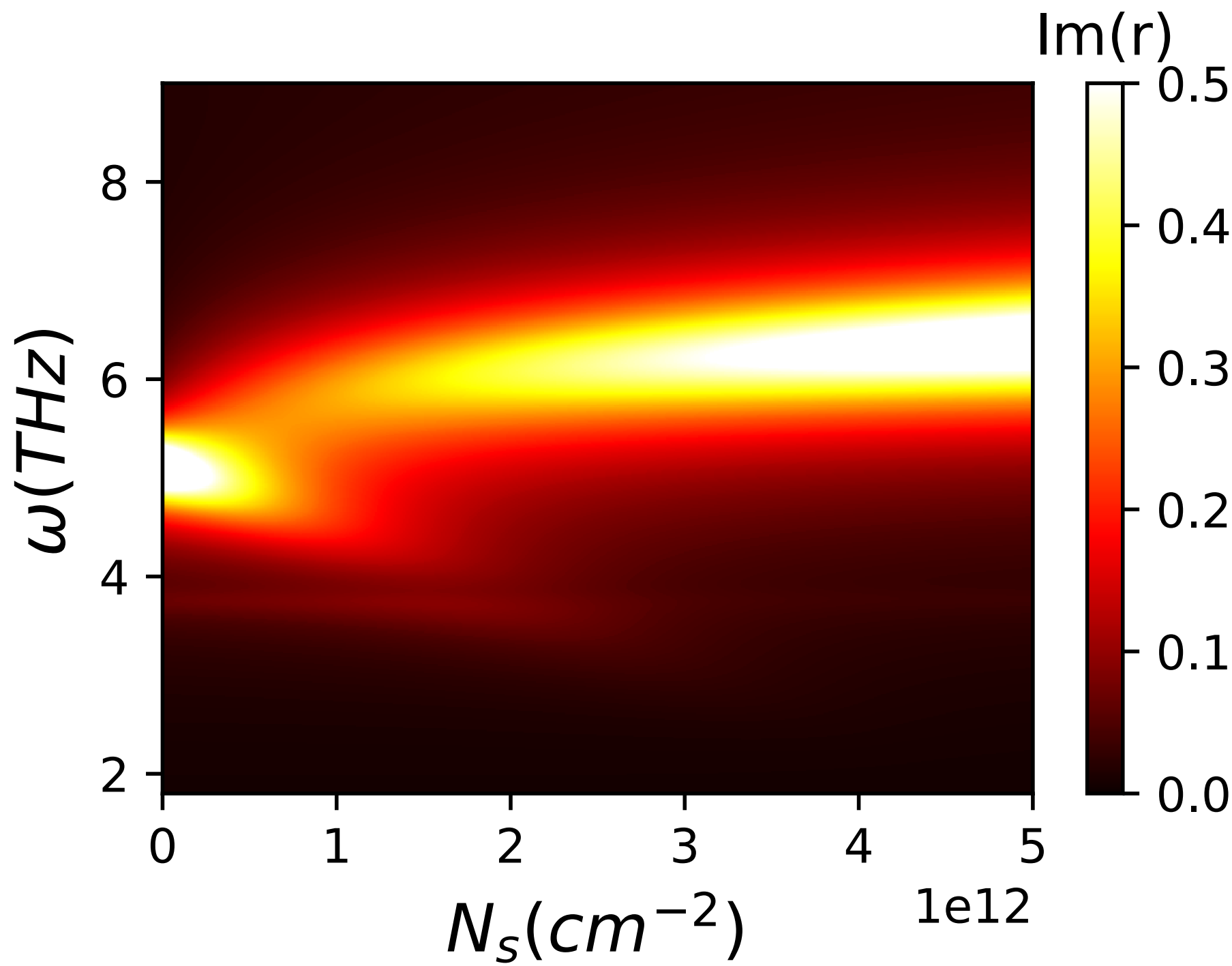}
 \caption{Dependence of the Dirac plasmon-phonon-ISBT polariton modes on the doping level in the QW. Calculations are performed at a specific wavevector $k_{x}=0.5 \times 10^{5} ~ cm^{-1}$ and with the thicknesses of the Bi$_{2}$Se$_{3}$ and the GaAs spacer respectively $d_{TI}=100~nm$ and $d_{sp}=10~nm$.}
  \label{FIG8}
\end{figure}

\begin{figure}[htb]
\centering
    \includegraphics[width=.5\textwidth]{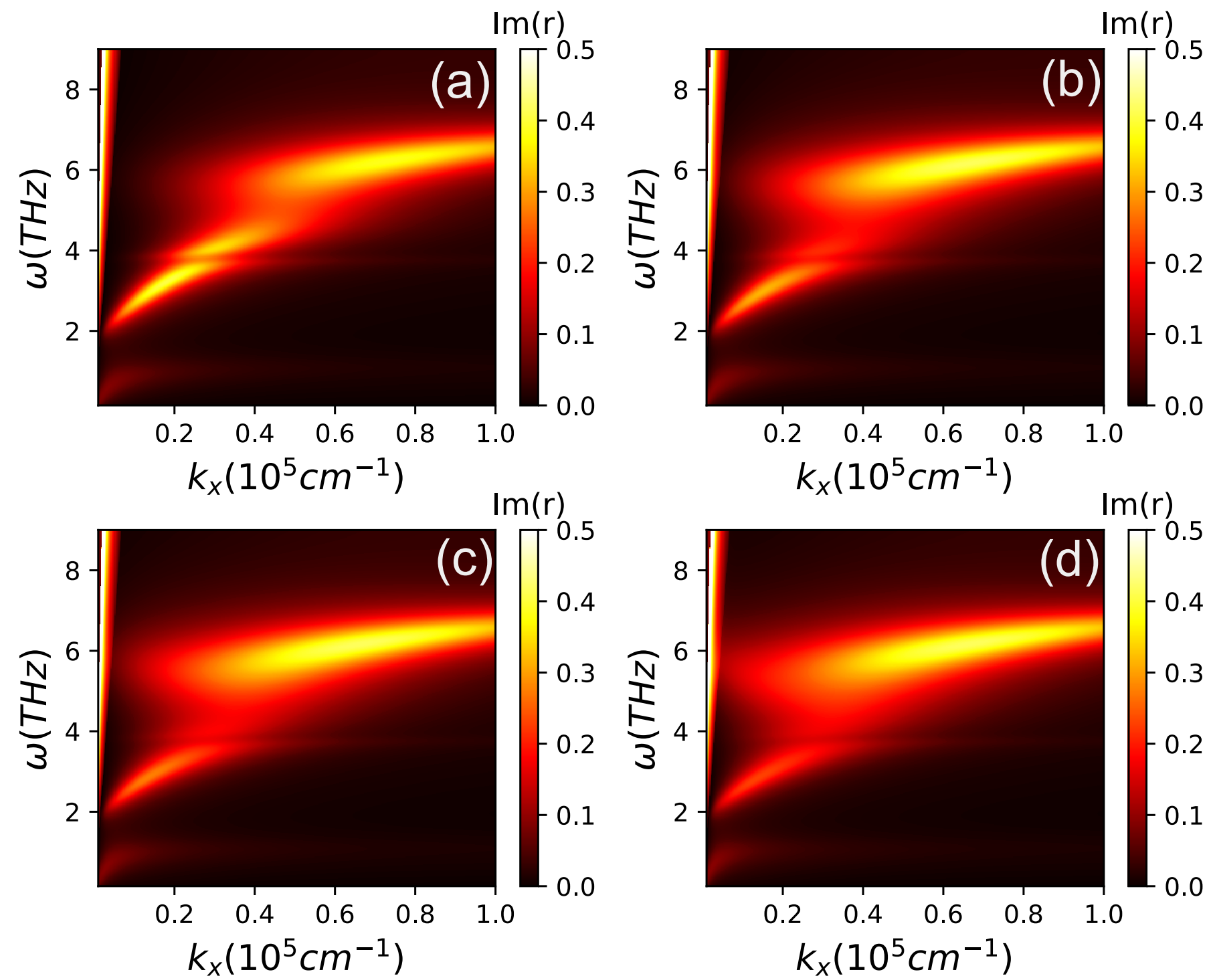}
 \caption{The dispersion relation of the Dirac plasmon-phonon-ISBT polarition modes calculated for (a) 1 AlGaAs QW, (b) 3 QWs, (c) 5 QWs, and (d) 10 QWs. Calculations were performed with $d_{TI}=100~nm$, $d_{sp}=10~nm$, and each single AlGaAs/GaAs QW having $d_{B}=100~nm$, $d_{w}=25~nm$, and $N_{s}=10^{12}~cm^{-2}$.}
  \label{FIG10}
\end{figure}

The dependence of polariton modes on the material excitations can be well described using the Dicke model in which, for a given photonic mode, the light-matter coupling is linearly-proportional to the square root of the total number of electrons within the resonator: $\Omega_{Rabi} \sim \sqrt{N_{e}}$ \cite{Torma2014}. We therefore consider how the plasmon-phonon-ISBT polariton dispersion relation depends on both the doping level in each QW ($N_{s}$) and the total number of QWs in the heterostructure. We first plot the polariton mode energy as a function of $N_{s}$ for a fixed value of wavevector $k_{x}=0.5 \times 10^{5}~cm^{-1}$. Other parameters are kept fixed at the values described in Sect.~\ref{method}, including the scattering loss rates that were varied in the previous section. As shown in Fig.~\ref{FIG8}, the polariton mode associated with coupling to the ISBT (at $\omega=6~THz$) does not appear when the doping is below $N_{s}\sim0.5\times 10^{12}~cm^{-2}$; only the plasmon-phonon polaritons in the TI (at $\omega\sim5~THz$) are observed for these low doping densities. As the doping increases, the number of carriers that can participate in the ISBT increases, and thus the strength of the coupling between the plasmon-phonon-polariton and the ISBT increases. This results in the emergence of the upper branch of the hybridized polariton at $\omega=6~THz$ for $N_{s}\gtrapprox0.5\times 10^{12}~cm^{-2}$. The intensity of this upper polariton branch continues to increase with doping density, without saturation over the range of doping densities considered here. We note, however, that the lower branches of the polariton become progressively fainter as $N_{s}$ increases, as can be seen in  Fig.~\ref{FIG8}. In  Fig.~\ref{FIG10} we consider the impact of increasing the number of QWs ($N_{QW}$) in the heterostructure.  Fig.~\ref{FIG10}(a) shows the Dirac plasmon-phonon-ISBT polaritons calculated with only $N_{QW}=1$ QW in the system; panels (b), (c), and (d) show the results calculated for $N_{QW}=3$, 5, and 10 QWs, respectively. Increasing the number of QWs increases the number of electrons participating in ISBT transitions and increases the splitting. This can be seen by observing that the upper polariton branch is clearly visible at at $k_{x}=0.2 \times 10^{5}~cm^{-1}$ when there are 10 QWs in the structure (Fig.~\ref{FIG10}d), but does not reach comparable intensity when there is 1 QW (Fig.~\ref{FIG10}a) until $k_{x}\gtrapprox0.4 \times 10^{5}~cm^{-1}$.  We note, however, that the increase in splitting is not linear with the number of QWs; the change between 5 and 10 QWs, for example, is nearly negligible, as can be seen by comparing Fig.~\ref{FIG10}(c) and (d). Because the strength of interactions between the DPP in the Bi$_{2}$Se$_{3}$ and the ISBT in a QW decrease as the distance between them increases, adding additional QWs that are relatively far removed from the Bi$_{2}$Se$_3$ layer has little impact. This stands in contrast to absorption measurements of ISBTs in QWs, which are typically performed in a transmission geometry where the observed absorption linearly increases with the number of identical QWs. These results tell us that an experimental measurement of the strong coupling modeled here would not require a sample with more than 3-5 QWs, which can significantly reduce sample growth time. 

\subsection{Effect of 2DHG at the Bi$_{2}$Se$_{3}$/GaAs(001) interface}

Finally, we consider the effects of the extra 2DHG at the interface between the Bi$_{2}$Se$_{3}$ and the GaAs(001) layers, which was discussed in Sect.~\ref{DFTpredict}. We note that we are not the first to consider the existence of a massive two dimensional gas of carriers at the interface between a TI and another material. For instance, Stauber {\em et al}.~added a massive two dimensional electron gas (2DEG) plasmon response to their analytical model in order to obtain a better fit to the experimental data they were modeling \cite{Stauber2017, Pietro2013}. 

\begin{figure}[htb]
\centering
    \includegraphics[width=.5\textwidth]{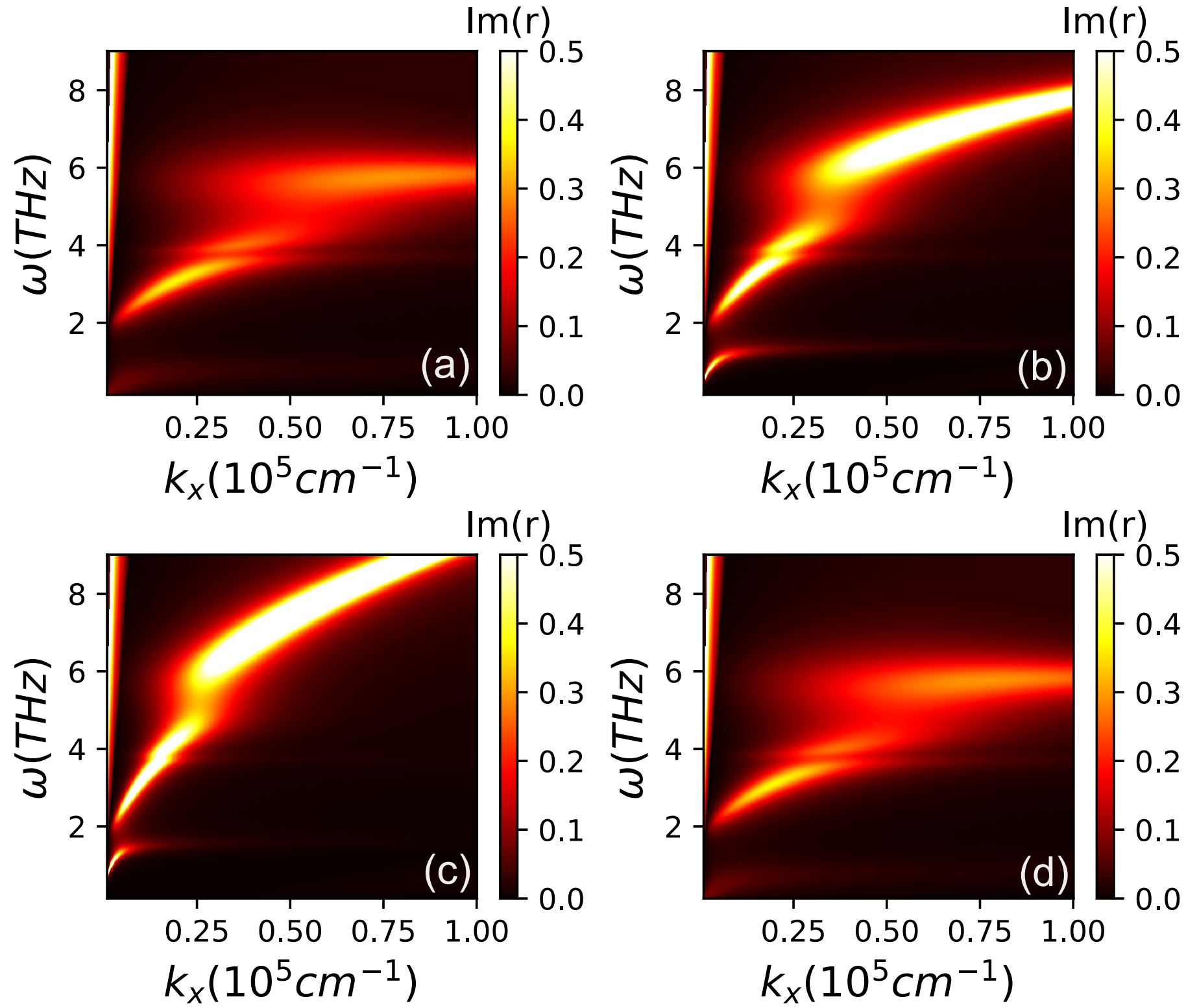}
 \caption{The dispersion relation of the Dirac plasmon-phonon-ISBT polariton modes for several possible values of the carrier concentration and scattering rate for holes in the 2DHG at the Bi$_{2}$Se$_{3}$/GaAs interface: (a) $N_{2DHG}=0,~\gamma_{2DHG}=0$, (b) $N_{2DHG}=5 \times 10^{13}~cm^{-2},~\gamma_{2DHG}=0$, (c) $N_{2DHG}= 10^{14}~cm^{-2},~\gamma_{2DHG}=0$, and (d) $N_{2DHG}=10^{14}~cm^{-2},~\gamma_{2DHG}=10~THz$ with the effective mass $m_{2DHG}^{*} \approx m_{0}$, where $m_{0}$ is the electron rest mass. Calculations performed with $d_{TI}=100~nm$, $d_{sp}=10~nm$, $d_{B}=100~nm$, $d_{w}=25~nm$, and $N_{s}=10^{12}~cm^{-2}$.}
  \label{FIG9}
\end{figure}

Here, our DFT calculations predict the existence of a 2DHG at the Bi$_{2}$Se$_{3}$/GaAs(001) interface. Because the surface states correlated with the 2DHG are described by rather flat bands, we can estimate that the interface 2DHG will have a carrier concentration significantly larger than that of the DPP in the bare Bi$_{2}$Se$_{3}$ layer, which is of order $10^{13}~cm^{-2}$. In Fig.~\ref{FIG9}(a)-(c) we plot the dispersion of the plasmon-phonon-ISBT polariton for 2DHG carrier concentrations of 0, $N_{2DHG}=5 \times 10^{13}~cm^{-2}$, and $N_{2DHG}=10^{14}~cm^{-2}$. All three of these calculations are done assuming that the interface between Bi$_{2}$Se$_{3}$ and GaAs is extremely clean so that the scattering rate of electrons at the interface is negligible ($\gamma_{2DHG}\sim0$). We observe that the dispersion of the plasmon-phonon-ISBT polariton simply shifts to higher frequency as $N_{2DHG}$ increases. This can be understood by looking back to Eq.~\ref{BiSedispersion}: the resonant frequency of the DPP in the Bi$_{2}$Se$_{3}$ layer depends on the square root of total carrier concentration, including the contribution from both surfaces of the Bi$_{2}$Se$_{3}$. Thus when the 2DHG adds more carriers at the interface, the DPP and, consequently, the hybridized excitation shift to higher frequencies. In Fig.~\ref{FIG9}(d) we compute what happens for a 2DHG carrier concentration of $N_{2DHG} = 10^{14}~cm^{-2}$ if the Bi$_{2}$Se$_{3}$/GaAs(001) interface were of poor quality, characterized by a scattering loss rate of $\gamma_{2DHG}=10~THz$. We see that the blue shift in the hybridized polariton frequency is completely suppressed when the scattering rate is larger. This result emphasize the requirement of sharp, defect-free interfaces when experimentally exploring the consequences of interface states in a hybrid material such as the one we model here. 

\section{Conclusions}
\label{conc}
We employ a scattering matrix approach to systematically investigate the coupling between THz excitations in the topological insulator Bi$_{2}$Se$_{3}$ and the ISBT transitions of III-V QWs. Our results show that spectral signatures of strong coupling, specifically hybridized plasmon-phonon-ISBT polaritons with splitting comparable to twice the FWHM linewidth of either hybridizing excitation, can emerge when (a) the thickness of the GaAs spacer layer separating the Bi$_{2}$Se$_{3}$ from the III-V QWs is sufficiently small and b) the quality of the materials is sufficiently high. We systematically analyze the dependence of the predicted spectra on parameters such as the number of QWs, the doping density of the QWs, the scattering loss rate in each materials, the doping of an interface 2DHG, and the phonons within the Bi$_{2}$Se$_{3}$. From this analysis we determine that the phonons in the Bi$_{2}$Se$_{3}$ play a critical role in the coupling process and that hybridized states could not be observed without these phonons. We also determine the structural and material quality parameters that must be achieved in order to experimentally explore the strong coupling regime in these hybrid material system. In particular, and as reported in Table \ref{table1}, approaching the strong coupling regime requires $d_{sp} \sim 10$ nm, $\gamma_{ISBT} \leq 1$ THz, and $N_{s}\gtrapprox0.5\times 10^{12}~cm^{-2}$. The strength of the coupling and the magnitude of the resulting splitting ($g$) increases from 0.7 THz to 1 THz as the number of QWs increases from 1 to 5, but does not increase beyond this point because the additional QWs are farther and farther from the interface with the TI. 

\begin{acknowledgments}
This research was primarily supported by NSF through the University of Delaware Materials Research Science and Engineering Center, DMR-2011824. The DFT calculations were performed at the the Extreme Science and Engineering Discovery Environment (XSEDE) supercomputer facilities, supported by National Science Foundation grant number ACI-1053575.
\end{acknowledgments}

\appendix

\section{Scattering and transfer matrix method}
\label{Smatrix}
\begin{figure}[h]
\centering
    \includegraphics[width=.49\textwidth]{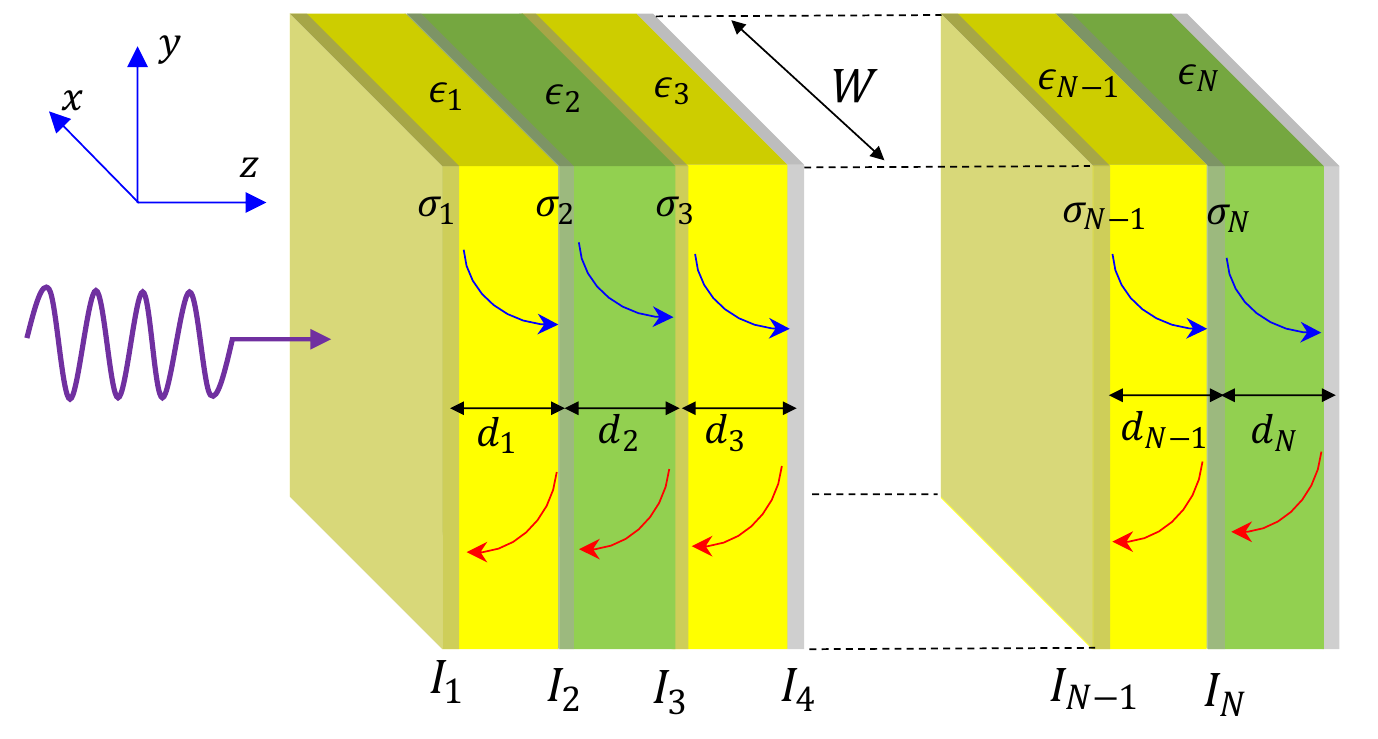}
 \caption{Schematic of a multilayer structure comprised of N constituent materials considered in this appendix. The growth direction of the structure is chosen along z-axis with an assumption that all layers have same ribbon's width $W$ along x-direction. The thickness, permittivity, and optical conductivity of the electron sheet in the $i^{th}$ layer are denoted by $d_{i}$, $\epsilon_{i}$ and $\sigma_{i}$, respectively whereas $I_{i}$ indicates the interface matrix at the $i^{th}$ interface.}
  \label{FIG4}
\end{figure}

In this appendix we describe the scattering and transfer matrix method for the general case of a multilayer structure comprised of N constituent materials with respective thickness $d_{i}$ of the $i^{th}-layer$ as depicted in Fig.~\ref{FIG4}. $\sigma_{i}$ represents the optical conductivity of the 2-dimensional electron sheet at the $i^{th}$ interface and $\epsilon_{i}$ is the permittivity tensor of the material sandwiched between the $i^{th}$ and $(i+1)^{th}$ interfaces. The work reported here does not consider magnetic materials, so we always set the magnetic permittivity $\mu=\mu_{0}$, which is the magnetic permeability of free space. 

Within each layer, the electric field $\boldsymbol{E}=\left(E_{x},E_{y},E_{z} \right)$ and the magnetic field $\boldsymbol{H}=\left(H_{x},H_{y},H_{z} \right)$ of a monochromatic electromagnetic wave propagating along the z direction take the general form:

\begin{align}
E_{x,i}&=\left( A_{x,i}e^{ik_{z,i}z } + B_{x,i}e^{-ik_{z,i}z}\right)e^{ik_{x,i}x}e^{-i\omega t} \label{EM1} \\
E_{y,i}&=\left( A_{y,i}e^{ik_{z,i}z}  + B_{y,i}e^{-ik_{z,i}z}\right)e^{ik_{x,i}x}e^{-i\omega t} \label{EM2} \\
E_{z,i}&=-\frac{\epsilon_{xx}k_{x}}{\epsilon_{zz}k_{z,i}}\left( A_{x,i}e^{ik_{z,i}z} -  B_{x,i}e^{-ik_{z,i}z} \right)e^{ik_{x,i}x}e^{-i\omega t} \label{EM3}
\end{align}
and 
\begin{align}
H_{x,i}&=-\frac{k_{z,i}}{\mu_{0}\omega}\left( A_{y,i}e^{ik_{z,i}z } - B_{y,i}e^{-ik_{z,i}z}\right)e^{ik_{x,i}x}e^{-i\omega t} \label{EM4} \\
H_{y,i}&= \frac{\epsilon_{0}\epsilon_{i}^{xx}\omega}{k_{z,i}}\left( A_{x,i}e^{ik_{z,i}z} - B_{x,i}e^{-ik_{z,i}z} \right)e^{ik_{x,i}x}e^{-i\omega t} \label{EM5} \\
H_{z,i}&=\frac{k_{x,i}}{\mu_{0}\omega}\left( A_{y,i}e^{ik_{z,i}z} -  B_{y,i}e^{-ik_{z,i}z} \right)e^{ik_{x,i}x}e^{-i\omega t} \label{EM6}
\end{align}
where $A_{(x,y),i}$ and $B_{(x,y),i}$ are the amplitudes of the x- and y- components of forward- and backward-propagating electromagnetic waves, respectively, $\omega$ is the frequency of the EM wave, $k_{x,i}$ and $k_{z,i}$ are the x- and z-components of the wavevector of the EM wave within the $i^{th}$ layer, x and z are the coordinate along the x- and z- directions. If we define $\hat{n}$ to be the unit vector perpendicular to the $i^{th}$ interface, $\boldsymbol{J}_{i}$ to be the in plane current, and $\rho_{i}$ to be the carrier density of the electron gas at the $i^{th}$ interface, we can then write down the boundary conditions for the electric and the magnetic fields at the $(i)^{th}$ interface: 
\begin{align}
\hat{n} \times \left( \boldsymbol{H}_{i+1} - \boldsymbol{H}_{i}\right)\vert_{z=z_{i}} &= \boldsymbol{J}_{i} \label{boundary1} \\
\hat{n} \times \left( \boldsymbol{E}_{i+1} - \boldsymbol{E}_{i}\right)\vert_{z=z_{i}} &= 0 \label{boundary2} \\
\hat{n}. \left( \boldsymbol{D}_{i+1} - \boldsymbol{D}_{i}\right)\vert_{z=z_{i}} &= \rho_{i} \label{boundary3} \\
\hat{n}. \left( \boldsymbol{B}_{i+1} - \boldsymbol{B}_{i}\right)\vert_{z=z_{i}} &= 0 \label{boundary4} \\
\boldsymbol{D}=\epsilon_{0}\epsilon \boldsymbol{E}, ~ ~ ~ ~ \boldsymbol{H} = \frac{1}{\mu_{0}}\boldsymbol{B}.
\end{align}

We now consider a scattering event taking place at the $i^{th}$ interface. Using the boundary conditions \ref{boundary1}-\ref{boundary4} we can compute the electric field amplitudes of the x- and y-polarized components of the left-propagating (A) and right-propagating (B) EM waves:
\begin{equation}
    \begin{pmatrix}
A_{x,i} \\
A_{y,i} \\
B_{x,i} \\
B_{y,i}
\end{pmatrix} = I_{i} \begin{pmatrix}
A_{x,i+1} \\
A_{y,i+1} \\
B_{x,i+1} \\
B_{y,i+1}
\end{pmatrix}
\end{equation}
with
\begin{widetext}
\begin{equation}
   I_{i} = \begin{bmatrix}
1+\frac{\epsilon_{i+1}^{xx}k_{z,i}}{\epsilon_{i}^{xx}k_{z,i+1}}+\frac{k_{z,i}\sigma_{i}^{xx}}{\epsilon_{0}\epsilon_{i}^{xx}\omega} &\frac{k_{z,i}\sigma_{i}^{xy}}{\epsilon_{0}\epsilon_{i}^{xx}\omega}  & 1-\frac{\epsilon_{i+1}^{xx}k_{z,i}}{\epsilon_{i}^{xx}k_{z,i+1}}+\frac{k_{z,i}\sigma_{i}^{xx}}{\epsilon_{0}\epsilon_{i}^{xx}\omega}  &\frac{k_{z,i}\sigma_{i}^{xy}}{\epsilon_{0}\epsilon_{i}^{xx}\omega}\\

\frac{\mu_{0}\omega\sigma_{i}^{yx}}{k_{z,i}} & 1+\frac{k_{z,i+1}}{k_{z,i}}+\frac{\mu_{0}\omega\sigma_{i}^{yy}}{k_{z,i}} & \frac{\mu_{0}\omega\sigma_{i}^{yx}}{k_{z,i}}   & 1-\frac{k_{z,i+1}}{k_{z,i}}+\frac{\mu_{0}\omega\sigma_{i}^{yy}}{k_{z,i}}\\

1-\frac{\epsilon_{i+1}^{xx}k_{z,i}}{\epsilon_{i}^{xx}k_{z,i+1}}-\frac{k_{z,i}\sigma_{i}^{xx}}{\epsilon_{0}\epsilon_{i}^{xx}\omega} &-\frac{k_{z,i}\sigma_{i}^{xy}}{\epsilon_{0}\epsilon_{i}^{xx}\omega} &1+\frac{\epsilon_{i+1}^{xx}k_{z,i}}{\epsilon_{i}^{xx}k_{z,i+1}}-\frac{k_{z,i}\sigma_{i}^{xx}}{\epsilon_{0}\epsilon_{i}^{xx}\omega}  &-\frac{k_{z,i}\sigma_{i}^{xy}}{\epsilon_{0}\epsilon_{i}^{xx}\omega}\\

-\frac{\mu_{0}\omega\sigma_{i}^{yx}}{k_{z,i}}  & 1-\frac{k_{z,i+1}}{k_{z,i}}-\frac{\mu_{0}\omega\sigma_{i}^{yy}}{k_{z,i}} & -\frac{\mu_{0}\omega\sigma_{i}^{yx}}{k_{z,i}}  & 1+\frac{k_{z,i+1}}{k_{z,i}}-\frac{\mu_{0}\omega\sigma_{i}^{yy}}{k_{z,i}}\\
\end{bmatrix}
\end{equation}
\end{widetext}
where
\begin{equation}
    k_{z,i}=\sqrt{\frac{\epsilon^{xx}_{i}\omega^{2}}{c^{2}}-\frac{\epsilon^{xx}_{i}}{\epsilon^{zz}_{i}}k_{x,i}^{2}}.
\end{equation}
The in-plane wave vector is given by
\begin{equation}
    k_{x,i}\left(i=1:N\right) 	\equiv q \approx \frac{\pi}{W}.
\end{equation}
where $W$ is the ribbon's width (see figure \ref{FIG4}). The tensor conductivity of two dimensional electron sheet at the $i^{th}$ interface is given by:
\begin{equation}
    \sigma_{i} =\begin{pmatrix} \sigma_{i}^{xx} & \sigma_{i}^{xy} \\
    \sigma_{i}^{yx} & \sigma_{i}^{yy}
    \end{pmatrix}.
\end{equation}
In the case when no external magnetic field is applied, which is the case here, one obtains:
\begin{align}
    \sigma_{i}^{xy}=\sigma_{i}^{yx} = 0 \\
    \sigma_{i}^{xx}=\sigma_{i}^{yy} = \sigma_{i}
\end{align}

Once we have the matrix that captures what happens at every interface we can construct a transfer matrix, $T$, for a structure comprised of N layers by computing:
\begin{equation}
    T = I_{1}P^{1}I_{2}P^{2} \dotsm P^{N-1}I_{N}
\end{equation}
which satisfies the relation:
\begin{equation}
    \begin{pmatrix}
    A_{x,1}\\
    A_{y,1}\\
    B_{x,1}\\
    B_{y,1}
    \end{pmatrix} = T \begin{pmatrix}
    A_{x,N+1}\\
    A_{y,N+1}\\
    B_{x,N+1}\\
    B_{y,N+1}
    \end{pmatrix}
\end{equation}
where $P^{i}$ is the propagation matrix:
\begin{equation}
    P^{i}=\begin{pmatrix}
    e^{ik_{z,i}d_{i}} &0 &0 &0\\
    0 &e^{ik_{z,i}d_{i}} &0 &0\\
    0 &0 &e^{-ik_{z,i}d_{i}} &0\\
    0 &0 &0 &e^{-ik_{z,i}d_{i}}\\
    \end{pmatrix}
\end{equation}
with $d_{i}$ the corresponding thickness of the $i^{th}$ layer. Note that the $T$ matrix can be written as a $2 \times 2$ matrix:
\begin{equation}
    T = \begin{bmatrix}
    T_{11} & T_{12} \\
    T_{21} & T_{22}
    \end{bmatrix}
\end{equation}
in which each element (e.g. $T_{ij}$ where $i,j=1,2$) is itself a $2 \times 2$ matrix.

In the same manner, we define a scattering matrix $S_{0,i}$ which connects the amplitudes of EM wave on the left side of 1$^{st}$  interface to those on the right side of $i^{th}$ interface:
\begin{equation}
    \begin{pmatrix}
    A_{x,i}\\
    A_{y,i}\\
    B_{x,0}\\
    B_{y,0}
    \end{pmatrix} = S_{0,i} \begin{pmatrix}
    A_{x,0}\\
    A_{y,0}\\
    B_{x,i}\\
    B_{y,i}
    \end{pmatrix}
\end{equation}
and a transfer matrix $IM_{i,i+1}$ which connects the amplitudes of EM wave on the right side of $i^{th}$ interface to those on the right side of $(i+1)^{th}$ interface:
\begin{equation}
    \begin{pmatrix}
    A_{x,i}\\
    A_{y,i}\\
    B_{x,i}\\
    B_{y,i}
    \end{pmatrix} = IM_{i,i+1} \begin{pmatrix}
    A_{x,i+1}\\
    A_{y,i+1}\\
    B_{x,i+1}\\
    B_{y,i+1}
    \end{pmatrix}
\end{equation}
then a scattering matrix $S_{0,i+1}=S_{0,i} \otimes IM_{i,i+1}$ that relates the outgoing and incoming states at the $1^{st}$ interface to the left side and the $(i+1)^{th}$ interface to the right side:
\begin{equation}
    \begin{pmatrix}
    A_{x,i+2}\\
    A_{y,i+2}\\
    B_{x,i}\\
    B_{y,i}
    \end{pmatrix} = S_{i,i+1} \begin{pmatrix}
    A_{x,i}\\
    A_{y,i}\\
    B_{x,i+2}\\
    B_{y,i+2}
    \end{pmatrix}
\end{equation}
will be obtained via recursive method \cite{To2019}:
\begin{widetext}
\begin{align}
    S_{0,i+1}^{11} &= \left[\mathcal{I} - \left(IM_{i,i+1}^{11} \right)^{-1}S_{0,i}^{12}IM_{i,i+1}^{21} \right]^{-1}\left(IM_{i,i+1}^{11} \right)^{-1}S_{0,i}^{11}  \\
    S_{0,i+1}^{12} &= \left[\mathcal{I} - \left(IM_{i,i+1}^{11} \right)^{-1}S_{0,i}^{12}IM_{i,i+1}^{21} \right]^{-1}\left(IM_{i,i+1}^{11} \right)^{-1}\left(S_{0,i}^{12}IM_{i,i+1}^{22}-IM_{i,i+1}^{12} \right)  \\
    S_{0,i+1}^{21} &=S_{0,i}^{22}IM_{i,i+1}^{21} S_{0,i+1}^{11} + S_{0,i}^{21}  \\
    S_{0,i+1}^{22}&=S_{0,i}^{22}IM_{i,i+1}^{21} S_{0,i+1}^{12} + S_{0,i}^{22}IM_{i,i+1}^{22} 
\end{align}
\end{widetext}
where $\mathcal{O}_{i,j}^{11}, ~ \mathcal{O}_{i,j}^{12}, ~ \mathcal{O}_{i,j}^{21}, ~ \mathcal{O}_{i,j}^{22}$ are $2\times2$ block elements of the matrix $\mathcal{O}_{i,j}~(\mathcal{O} \equiv S, IM)$  and $\mathcal{I}$ indicates the identity matrix. Overall, one can construct the total scattering matrix $S=S_{0,N}=S_{0,1} \otimes IM_{1,2} \otimes ... \otimes IM_{N-1,N}$ which links the amplitudes of electromagnetic waves in the outer and inner region:
\begin{equation}
    \begin{pmatrix}
    A_{x,N+1}\\
    A_{y,N+1}\\
    B_{x,1}\\
    B_{y,1}
    \end{pmatrix} = S \begin{pmatrix}
    A_{x,1}\\
    A_{y,1}\\
    B_{x,N+1}\\
    B_{y,N+1}
    \end{pmatrix}
\end{equation}
leading to the well-known relationship between $S$ and $T$:
\begin{equation}
    S = \begin{bmatrix}
    T_{11}^{-1} & -T_{11}^{-1}T_{12}\\
    T_{21}T_{11}^{-1} & T_{22}-T_{21}T_{11}^{-1}T_{12}
    \end{bmatrix} = \begin{bmatrix}
    t & r' \\
    r & t'
    \end{bmatrix}
\end{equation}
where $t,t'$ and $r,r'$ are each $2 \times 2$ matrices indicating the transmission and reflection coefficients. Here $t$ and $r$ are respective the transmission and reflection associating with incident wave propagating along +z direction whereas $t'$ and $r'$ correspond to the incident wave propagating along -z direction. This formulation allows for an explicit picture of the reflection and transmission coefficients for the TE and TM polarized EM waves. To illustrate how this formulation allows us to extract the optical properties of multilayer structures, consider the case in which a TM polarized wave is incident on the structure in Fig.~\ref{FIG4}. At the $1^{st}$ interface there are incident and reflected waves which in general have both TE- and TM-polarized components. If we write the reflection matrix $r$ as
\begin{equation}
    r = \begin{pmatrix}
    r_{11} & r_{12} \\
    r_{21} & r_{22}
    \end{pmatrix}
\end{equation}
then the reflection coefficient associated with the TE reflected wave is $r_{TE}=r_{21}$ and the reflection coefficient associated with the TM reflected wave is $r_{TM}=r_{11}$. The reflectances of the TE and TM reflected waves are then given by:
\begin{align}
    R_{TM} &= r_{TM}^{2} \\
    R_{TE} &= \left[ \left(\frac{k_{z,1}}{\mu_{0}\omega}\right)^{2} + \left( \frac{q}{\mu_{0}\omega} \right)^{2} \right]r_{TE}^{2}
\end{align}
Analogously, in the case of a TE polarized wave incident on the structure the reflection coefficients associated with the TE and TM reflected waves are $r_{TE}=r_{22}$ and $r_{TM}=r_{12}$, respectively.  The reflectances of the TE and TM reflected waves are then:
\begin{align}
    R_{TM} &= \left[ \left( \frac{k_{z,1}}{\epsilon_{0}\epsilon_{1}\omega} \right)^{2}  + \left(\frac{q}{\epsilon_{0}\epsilon_{1}\omega} \right)^{2}\right]r_{TM}^{2} \\
    R_{TE} &= r_{TE}^{2}
\end{align}

Using these equations, we can compute the transmission and reflection coefficients for any multilayer structure. The imaginary part of the reflection coefficient, $Im(r)$, is proportional to the losses in the system. The presence of loss in the reflectance spectrum indicates that the incident EM wave has generated an excitation that is carrying energy away laterally, i.~e.~propagating in the x- or y- direction rather than transmitting or reflecting in the +z or -z directions, respectively. The frequency dependence of such loss thus generates the dispersion curves for the hybridized excitations in the coupled system, which is what we wish to study.

\section{Structural parameters}\label{parameters}
\begin{widetext}
\begin{table*}[h!]
\caption{Structural parameters used for the figures in this work}
\tiny
\begin{tabular}{ |c||c|c|c|c|c|c|c|c|c|c|c|c|c|}
\hline
Figure & $d_{TI}$ (nm) & $d_{sp}$ (nm)  &$d_{B}$ (nm) & $d_W$  (nm)&$N_{s}~(cm^{-2})$ &$N_{2DHG}~(cm^{-2})$ &$N_{QW}$ &$\gamma_{plasmon}$ (THz) &$\gamma_{TI}$ (THz) &$\gamma_{ISBT}$(THz) &$\gamma_{2DHG}$ (THz) &g (THz) \\
\hline
\hline
5(a) &100 &500  &100 &25 &$1\times 10^{12}$ &0 &1 & 3 &$\alpha=0.53$, $\beta=0.3$ &1 &0 &0 \\
5(b) &100 &10  &100 &25 &$1\times 10^{12}$ &0 &1 & 3 &$\alpha=0.53$, $\beta=0.3$ &1 &0 &0.7 \\
6(a) &100 & 10  &100 &25 &$1\times 10^{12}$ &0 &1 & 3 &$\alpha=0.53$, $\beta=0.3$ &1 &0 &0 \\
6(b) &100 & 10  &100 &25 &$1\times 10^{12}$ &0 &1 & 3 &$\alpha=0.53$, $\beta=0.3$ &1 &0 &0 \\
6(c) &100 & 10  &100 &25 &$1\times 10^{12}$ &0 &1 & 3 &$\alpha=0.53$, $\beta=0.3$ &1 &0 &0.7 \\
6(d) &100 & 10  &100 &25 &$1\times 10^{12}$ &0 &1 & 3 &$\alpha=0.53$, $\beta=0.3$ &1 &0 &0.7 \\
8(a) &$0-200$ & 10  &100 &25 &$1\times 10^{12}$ &0 &1 & 3 &$\alpha=0.53$, $\beta=0.3$ &1 &0 &0.7 \\
8(b) &100 & $0-300$  &100 &25 &$1\times 10^{12}$ &0 &1 & 3 &$\alpha=0.53$, $\beta=0.3$ &1 &0 &$0.7-0$ \\
9(a) &100 & 10  &100 &25 &$1.5\times 10^{12}$ &0 &1 & 3 &$\alpha=0-2$, $\beta=0-2$ &1 &0 &0.7 \\
9(b) &100 & 10  &100 &25 &$1.5\times 10^{12}$ &0 &1 & 3 &$\alpha=0.53$, $\beta=0.3$  &$0-2$ &0 &$1-0$ \\
10 &100 & 10  &100 &25 &$0-5\times 10^{12}$ &0 &1 & 3 &$\alpha=0.53$, $\beta=0.3$  &1 &0 &$0-2$ \\
11(a) &100 & 10  &100 &25 &$1\times 10^{12}$ &0 &1 & 3 &$\alpha=0.53$, $\beta=0.3$  &1 &0 &0.7 \\
11(b) &100 & 10  &100 &25 &$1\times 10^{12}$ &0 &3 & 3 &$\alpha=0.53$, $\beta=0.3$  &1 &0 &0.8 \\
11(c) &100 & 10  &100 &25 &$1\times 10^{12}$ &0 &5 & 3 &$\alpha=0.53$, $\beta=0.3$  &1 &0 &1 \\
11(d) &100 & 10  &100 &25 &$1\times 10^{12}$ &0 &10 & 3 &$\alpha=0.53$, $\beta=0.3$  &1 &0 &1 \\
12(a) &100 & 10  &100 &25 &$1\times 10^{12}$ &0 &1 & 3 &$\alpha=0.53$, $\beta=0.3$  &1 &0 &0.7 \\
12(b) &100 & 10  &100 &25 &$1\times 10^{12}$ &$5\times10^{13}$ &1 & 3 &$\alpha=0.53$, $\beta=0.3$  &1 &0 &0.7 \\
12(c) &100 & 10  &100 &25 &$1\times 10^{12}$  &$10^{14}$ &1 & 3 &$\alpha=0.53$, $\beta=0.3$  &1 &0 &0.7 \\
12(d) &100 & 10  &100 &25 &$1\times 10^{12}$ &$10^{14}$ &1 & 3 &$\alpha=0.53$, $\beta=0.3$  &1 &10 &0.7 \\
\hline
\hline
\end{tabular}
\label{table1}
\end{table*}
\end{widetext}


\bibliography{Ref}

\end{document}